\documentclass{article}
\PassOptionsToPackage{numbers, compress}{natbib}
\usepackage[preprint]{neurips_2025}
\usepackage{multirow}

\usepackage[utf8]{inputenc} 
\usepackage[T1]{fontenc}    
\usepackage{hyperref}       
\usepackage{url}            
\usepackage{booktabs}       
\usepackage{amsfonts}       
\usepackage{nicefrac}       
\usepackage{microtype}      
\usepackage{xcolor}         
\usepackage{subcaption}
\usepackage{amsmath}
\usepackage{lscape}
\usepackage{rotating}
\usepackage{amssymb}
\usepackage{graphicx}
\usepackage{adjustbox}
\usepackage{booktabs}
\usepackage{physics}
\usepackage{makecell}
\usepackage{enumerate}
\usepackage{placeins}
\usepackage{enumitem}
\usepackage[normalem]{ulem}
\usepackage{colortbl}
\usepackage{authblk}
\usepackage{pdflscape} 
\usepackage{tabularx}  
\DeclareUnicodeCharacter{FF0C}{}
\usepackage{array}

\title{Unpaired Image-to-Image Translation for Segmentation and Signal Unmixing}

\author{%
  Nikola Andrejic, Milica Spasic, Igor Mihajlovic, Petra Milosavljevic, Djordje Pavlovic, Filip Milisavljevic, Uros Milivojevic, Danilo Delibasic, Ivana Mikic, Sinisa Todorovic \\
 \texttt{ivana@diffine.com}\\ Diffine LLC, 845 15th St Floor 1, San Diego, CA 92104, U.S.A. \\
}

\begin{document}

\maketitle

\begin{abstract}
This work introduces Ui2i, a novel model for unpaired image-to-image translation, trained on content-wise unpaired datasets to enable style transfer across domains while preserving content. Building on CycleGAN, Ui2i incorporates key modifications to better disentangle content and style features, and preserve content integrity. Specifically, Ui2i employs U-Net-based generators with skip connections to propagate localized shallow features deep into the generator. Ui2i removes feature-based normalization layers from all modules and replaces them with approximate bidirectional spectral normalization---a parameter-based alternative that enhances training stability.  To further support content preservation, channel and spatial attention mechanisms are integrated into the generators. Training is facilitated through image scale augmentation.  Evaluation on two biomedical tasks---domain adaptation for nuclear segmentation in immunohistochemistry (IHC) images  and unmixing of biological structures superimposed  in single-channel immunofluorescence (IF)  images---demonstrates Ui2i's ability to preserve content fidelity in settings that demand more accurate structural preservation than typical translation tasks. To the best of our knowledge, Ui2i is the first approach capable of separating superimposed signals in IF images using real, unpaired training data.
\end{abstract}


\section{Introduction}
\label{sec:intro}


In this paper, we introduce a novel unpaired image-to-image translation model, Ui2i, and demonstrate its effectiveness in two key applications: domain adaptation for nuclear segmentation and the separation of  the signal coming from different biological structures that are superimposed in single-channel immunofluorescence images. 

In digital pathology, spatial proteomics, and other microscopy applications, cellular and tissue structures are visualized through a range of staining techniques, such as hematoxylin and eosin (H\&E), immunohistochemistry (IHC), and immunofluorescence (IF). Due to the complexity and variability of biological structures captured by these staining methods, segmentation models \cite{stardist, instanseg, SAM4MIS} trained on one domain (e.g., H\&E) often fail to generalize effectively to others (e.g., IHC or IF). These models typically require fine-tuning not only to specific staining domain but also to each new dataset, given substantial variations in staining protocols and imaging conditions \cite{SAM4MIS}.  In practice, however, annotated training data for such fine-tuning is often scarce. One way to address this challenge is to use image-to-image (i2i) translation, which transforms new images to resemble the domain on which the model was originally trained, thereby improving cross-domain generalization. 

Beyond this application, one of our main contributions is to propose unpaired i2i translation as a novel framework for addressing another distinct challenge arising in multiplexed immunofluorescence (mIF) imaging. mIF enables simultaneous visualization of many molecular markers in a tissue sample, capturing the spatial complexity of biological structures. Increasing the number of detectable markers enhances the ability to study diverse cell types, interactions, and states. Physical constraints, such as spectral overlap, limit the number of markers that can be captured simultaneously. We apply Ui2i to a single-channel IF image to unmix multiple superimposed biological structures  into corresponding multi-channel representations, each isolating a distinct substructure of interest. Our computational marker separation by translation is particularly valuable, as it effectively doubles the capacity of mIF experiments by enabling a single fluorophore to label two different biomarkers instead of just one. By disentangling overlapping signals into well-separated channels, our approach reduces reliance on additional fluorophores and imaging rounds, thereby mitigating key limitations of current multiplexing technologies.



Since biological images are used for precise quantitative evaluation, any translation system must preserve image content---such as nuclear shapes and boundaries, cellular morphology, and spatial relationships between substructures---with the highest possible fidelity. This requirement is significantly more stringent than in many other applications of i2i translation, such as those involving natural scenes, where moderate distortions, hallucinated objects, or minor artifacts in the translated images are often tolerable and do not affect the end task. Unlike existing i2i translation methods \cite{wu2022single, CycleGAN, Long2015ICML}, our Ui2i is specifically designed to meet this stringent requirement.

Building on CycleGAN \cite{CycleGAN}, UI2I introduces key architectural and training improvements for bidirectional image-to-image translation between domains A and B. Ui2i is trained under the unpaired setting, where the two training datasets are unpaired in the sense that they depict the same type of biological structures but not the same biological samples.  Unlike CycleGAN’s ResNet-based generators, UI2I employs UNet-like generators with skip connections to propagate localized, shallow features to deeper layers, aiding fine-grained structure preservation. UI2I further departs from conventional i2i models by removing feature normalization layers and replacing them with approximate bidirectional spectral normalization---a parameter-based alternative that stabilizes training and enforces consistent representations of local structures (e.g., nuclei) regardless of global context. This design reduces translation artifacts such as droplet-like distortions and false positives typically caused by context-dependent feature normalization. To further enhance content preservation, UI2I integrates spatial and channel attention into the encoder modules. Finally, training is regularized with image scale augmentation, promoting scale-invariant feature learning.

%
%
Results show that Ui2i preserves image content fidelity to a significantly greater extent than existing i2i translation methods.

The rest of the paper: 
Sec.~\ref{sec:review} reviews related work. Sec.~\ref{sec:content_preservation} and \ref{sec:spectral_normalization} motivate and specify spectral normalization, Sec.~\ref{sec:approach} and \ref{sec:learning} present main components of Ui2i and its training.  Sec.~\ref{sec:segmentation} and \ref{sec:demultiplexing} report Ui2i's performance for domain adaptation in nuclear segmentation and unmixing in IF images.

\section{Related Work}
\label{sec:review}

Image-to-image (i2i) translation aims to map an image from one domain to another while preserving semantic content. Techniques in this field span several model families, including generative adversarial networks (GANs) \cite{GAN}, variational autoencoders (VAEs) \cite{Grosskopf2025}, and diffusion models \cite{xia2023diffi2iefficientdiffusionmodel}.  A key distinction among methods is whether they require \textit{paired} datasets (e.g., Pix2Pix \cite{Isola17}) or support \textit{unpaired} training (e.g., CycleGAN \cite{CycleGAN}), which is critical in biomedical imaging where pixel-aligned pairs are often unavailable. 

CycleGAN introduced a cycle consistency loss to enable unpaired translation, while subsequent models such as MUNIT \cite{Huang2018ECCV} and DRIT++ \cite{Lee2019DRITpp} achieve multimodal generation through content-style disentanglement. More recently, diffusion-based approaches such as StainDiff \cite{shen2023staindiff} have been applied to stain translation.

Image-to-image translation is widely used in digital pathology for tasks such as stain transfer \cite{Zingman2023}, stain normalization \cite{HOQUE2024101997}, and virtual staining \cite{LATONEN20241177}. It also supports downstream tasks such as cell and nuclear segmentation, as well as region segmentation \cite{kapil2019dasganjointdomain}.

Diversity of staining protocols and scanner characteristics in histopathology make domain adaptation a critical step, expanding the applicability of models trained for downstream tasks such as nuclear segmentation. Several studies reported using CycleGAN \cite{CycleGAN} to translate images to a domain of a pre-trained model \cite{Trullo2022, Liu2020}, as well as custom GAN-based architectures for unsupervised domain adaptation for the same purpose \cite{Haq2023}. 

As explained earlier, maximizing the number of fluorescent markers imaged in the same biological sample is of great value in biomedical research. Recent advances combine experimental innovations \cite{Zimmermann2003, Zimmermann2014, Datta2020, Lin2015, Goltsev2018} with computational techniques \cite{Lee2022, Choi2024, Tian2022} to overcome traditional multiplexing limits.  A largely unexplored direction is the use of deep learning to separate markers imaged with the same fluorophore - a capability that expands multiplexing in a way orthogonal to spectral, temporal, or chemical strategies and can be combined with them. To our knowledge, only one group has attempted this, using paired data: $\mu$Split \cite{ashesh2025indisplit, Ashesh2025.02.10.637323}  trains variational autoencoders on merged images created by summing individual marker channels or co-imaging spectrally distinct fluorophores. However, these combinations do not capture fully realistic appearance of co-labeled markers. To our knowledge, our method is the first to learn this separation from real, unpaired data. By training an unpaired image-to-image translation model between real single-channel and two-channel multiplexed images, we disentangle spectrally identical markers directly from experimental data.

\section{The Challenge of Content Preservation in Image-to-Image Translation}
\label{sec:content_preservation}

This work addresses unpaired bidirectional i2i translation \cite{lee2025, Yang24, xie2023unpaired, agrawal2020}. Most existing  GAN-based methods\cite{GAN} involve two generators, $G_{AB}$, and $G_{BA}$.  $G_{AB}$ translates real images from domain $A$ to $B$, $y'=G_{AB}(x)$, $x\in A$, while $G_{BA}$ performs the reverse mapping from domain $B$ to $A$, $x'=G_{BA}(y)$, $y\in B$. The resulting translations $x'$ and $y'$ are referred to as fake images. The generators are also used for identity translations---applying each generator to inputs from its target domain, $\hat{y}=G_{AB}(y)$, $y\in B$, and $\hat{x}=G_{BA}(x)$, $x\in A$. The outputs of identity translations are referred to as identity images. The generators are trained  jointly with a domain discriminator and a content discriminator through adversarial losses in a min-max optimization framework.

If the two translation directions remain uncoupled, the generators may map all inputs to a single output, resulting in mode collapse. To mitigate this, i2i translation methods commonly incorporate a cycle consistency loss \cite{CycleGAN}, $\mathcal{L}_\text{cyc}$, that couples the two directions of translation. For images $x\in A$ and $y\in B$, this loss is defined as 
%
%
%
 $   \mathcal{L}_\text{cyc} = \abs{x - G_{BA}(G_{AB}(x))} + \abs{y - G_{AB}(G_{BA}(y))}$.
%
Cycle consistency loss has been shown to encourage the disentanglement of features into those shared across domains---referred to as content features---and those specific to each domain---style features \cite{CycleGAN}. While enabling feature disentanglement,  $\mathcal{L}_\text{cyc}$ does not necessarily facilitate the preservation of content features during translation, as the following example illustrates. Given a pair of generators that perfectly disentangle content and style, $(\hat{G}_{AB}, \hat{G}_{BA})$, for any invertible transformations $T_A$ and $T_B$ acting in domains $A$ and $B$, respectively, the pairs $\left(G_{AB}', G_{BA}'\right) = (T_B \hat{G}_{AB}, \hat{G}_{BA} T_B^{-1})$ and $(G_{AB}'', G_{BA}'') = (\hat{G}_{AB} T_A^{-1}, T_A \hat{G}_{BA})$ are also cycle-consistent. This implies that $\mathcal{L}_\text{cyc}$ alone cannot uniquely determine a disentangled representation---i.e., there exist infinitely many generator pairs that satisfy cycle consistency while encoding different mappings of content and style.

\begin{figure*}[ht]
\centering
\begin{minipage}[c]{0.19\textwidth}
    \centering
    \includegraphics[width=\textwidth]{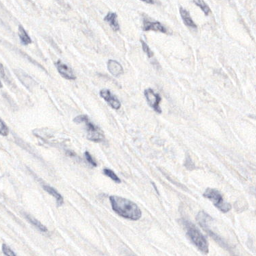}

    {\footnotesize (a) Input IHC image}
\end{minipage}\hfill  %
\begin{minipage}[c]{0.385\textwidth}
    \centering
            \includegraphics[width=.495\textwidth]{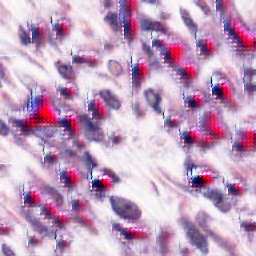}%
            \hfill
            \includegraphics[width=0.495\textwidth]{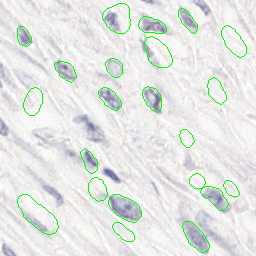}

     {\footnotesize (b) Translation by CycleGAN \cite{CycleGAN}}      
        \end{minipage}\hfill  %
        \begin{minipage}[c]{0.385\textwidth}
            \centering
            \includegraphics[width=0.495\textwidth]{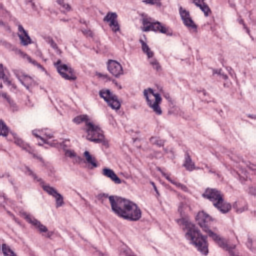}%
            \hfill
            \includegraphics[width=0.495\textwidth]{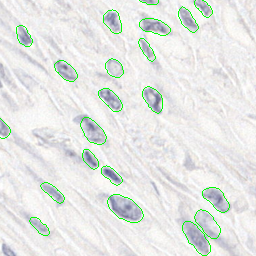}

    {\footnotesize (c) Translation by our Ui2i model}
        \end{minipage}
    \caption{Artifacts in image-to-image translation from the IHC to H\&E domain. (a) Input IHC image. (b) CycleGAN-translated H\&E image (b-left) shows numerous artifacts giving rise to false positives in the corresponding StarDist segmentation (b-right). (b) Ui2i-translated H\&E image (c-left) contains significantly fewer artifacts leading to improved StarDist segmentation (c-right). In both (b) and (c), StarDist \cite{stardist} is applied to the translated H\&E image, and the nuclear segmentation results (marked green) overlaid on the original input image for clarity.}
    \label{fig:blobs}
\end{figure*}

As a result of this uncertainty, translated images may exhibit undesirable artifacts such as shape distortions, hallucinated objects, and blob-like artifacts in the background, as illustrated in Fig.~\ref{fig:blobs}. In biomedical applications---especially for nuclear segmentation---these artifacts are very problematic, since blob-like hallucinated shapes can be mistaken for real nuclei, leading to many false positives, as shown in Fig.~\ref{fig:blobs}. The problem of blob-like artifacts is especially severe when instance normalization \cite{instancenorm} is used, as noticed in \cite{stylegan2}, where this layer is replaced by a parameter based $l_2$ normalization of weights. In Appendix, we present a toy example demonstrating that instance normalization causes localized objects in an image to have significantly different signal responses depending on the global context, whereas parameter-based normalizations remain resistant to this issue. Other feature-based normalization layers (such as batch normalization \cite{batchnorm} with large batch size, virtual batch normalization \cite{salimans2016}, group normalization \cite{groupnorm}, etc.) might be less susceptible to this issue, but they have their own drawbacks such as memory and computational overhead. Following  \cite{stylegan2}, we drop the feature-based normalization altogether and instead normalize only network parameters using bidirectional spectral normalization \cite{bsn, miyatospectral}, as detailed in the following section.


\section{Approximate Bidirectional Spectral Normalization}
\label{sec:spectral_normalization}
Spectral normalization (SN) of network parameters has been shown to improve the stability of GAN training \cite{miyatospectral} and prevent blob-like artifacts \cite{stylegan2}. In SN,  each layer's weight tensor ${\bf w}$ is divided by its spectral norm before being applied to the input. For convolutional layers,  ${\bf w}$ is first reshaped  to the forward reshape $W^{\text{FW}}$ by flattening the spatial and input channel indices. To avoid the high computational cost of exact SN, the spectral norm is approximated using a single iteration of the classical power method. Bidirectional Spectral Normalization (BSN) \cite{bsn} extends SN to better suit convolutional layers by addressing both the forward and backward flows of information. In addition to the standard forward reshape $W^{\text{FW}}$, BSN also introduces the backward reshape $W^{\text{BW}}$, which flattens the output channel and spatial dimensions of $w$. BSN normalizes ${\bf w}$ using the average of the spectral norms of $W^{\text{FW}}$ and $W^{\text{BW}}$, each approximated with a single iteration of the power method. 
Intuitively, $W^{\text{FW}}$ governs the flow of information during the forward pass, while $W^{\text{BW}}$ influences the propagation of gradients during backpropagation. 
By normalizing both $W^{\text{FW}}$ and $W^{\text{BW}}$, BSN regulates the Lipschitz continuity of the layer in both directions, improving training stability \cite{tensnorm, bsn}. 

As discussed in Sec. \ref{sec:content_preservation}, our Ui2i model removes all feature-based normalization layers and adopts BSN instead. For an efficient BSN approximation, we employ the following tight, differentiable lower bound on the spectral norm \cite{lowerbound}. Given the $m\times n$ forward (or backward) reshape matrix $W=(w_1,\dots,w_n)$, where $w_j$ are columns of $W$, its spectral norm $\|W\|$ can be lower-bounded with
\begin{equation}
    \|W\|\ge \sigma(W)=\frac{\left\|\left(\sum_{j=1}^n w_j w_j^\top  \right)r\right\|}{\|(w_1^\top r,\dots, w_n^\top r)^\top\|} ,
    \label{eq:bound}
\end{equation}
assuming the row sum vector $r = (r_1,\dots,r_m)^\top$ of the reshape matrix $W$ is nonzero. To normalize each layer's weight tensor ${\bf w}$, we use the root-mean-square (RMS) of the forward and backward norms, $\sigma_{\text{rms}}$, in place of their arithmetic mean, where   $\sigma_{\text{rms}}$ is computed as
\begin{equation}
\sigma_{\text{rms}}=\sqrt{\frac{\sigma^2(W^{\text{FW}}) + \sigma^2(W^{\text{BW}})}{2}}.
      \label{eq:rms}
\end{equation}
This combination---lower-bound estimation plus RMS aggregation---maintains training stability, accelerates convergence, and is straightforward to implement.

\section{Our Ui2i Model}
\label{sec:approach}
Building on CycleGAN \cite{CycleGAN}, our Ui2i model consists of: two generators, $G_{AB}$ and $G_{BA}$, one domain discriminator for classifying the outputs from the two generators, and another content discriminator for classifying deep features extracted by each generator. Ui2i incorporates several more extensions aimed at better preserving spatially localized content features compared to CycleGAN. 

\begin{figure}[ht]
    \centering
    \includegraphics[width=0.9\textwidth]{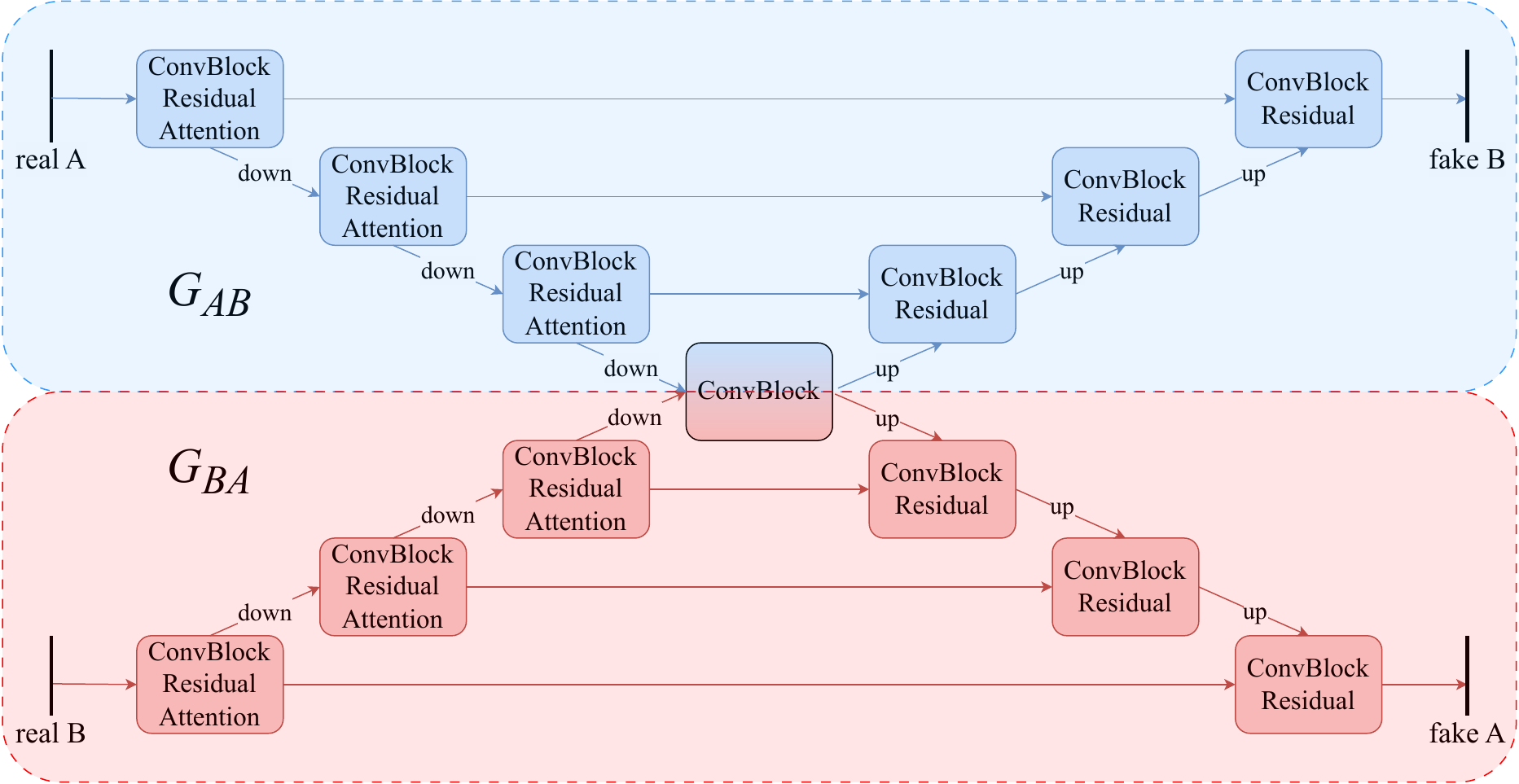}
    \caption{Our Ui2i comprises two generators, $G_{AB}$ and $G_{BA}$, that use a UNet-like architecture featuring skip connections and self-attention in certain encoder blocks (marked with ``Attention'') to better preserve spatial content. ``Residual'' denotes the usage of residual connections.}
    \label{fig:generator}
\end{figure}

Instead of using the original ResNet-based architecture, we adopt a UNet-like generator, as illustrated in Fig.~\ref{fig:generator}, which leverages the skip connections to better propagate fine-grained spatial information throughout the network. Moreover, the encoder blocks in Ui2i are equipped with additional self-attention mechanisms to refine feature representations and thereby enhance content preservation during translation.
The two generators, $G_{AB}$ and $G_{BA}$, share weights at the bottleneck block. Importantly, Ui2i eliminates feature-based normalization layers across all blocks and layers,  as motivated in Sec.~\ref{sec:content_preservation}. Instead, it applies bidirectional spectral normalization to weights of all layers except the generators' final layer and the final layer of the self-attention modules.

Rather than employing two separate domain discriminators, Ui2i adopts a single stacked domain discriminator \cite{stackedgan}, in which the input and translated images are concatenated along the channel dimension and classified into two (real and fake) or three (real, fake and identity) classes, depending on whether or not the identity loss is used. This design reduces overfitting and improves computational efficiency compared to using separate discriminators. Following \cite{Lee2019DRITpp}, Ui2i incorporates an additional content discriminator that classifies the domain of features extracted from the bottleneck block into two classes---domain A or domain B, encouraging these representations to be domain-invariant and thereby promoting content preservation. Below, we provide a detailed overview of the main components of the Ui2i architecture.

\begin{figure*}[ht]
    \centering
    \includegraphics[width=0.8\textwidth]{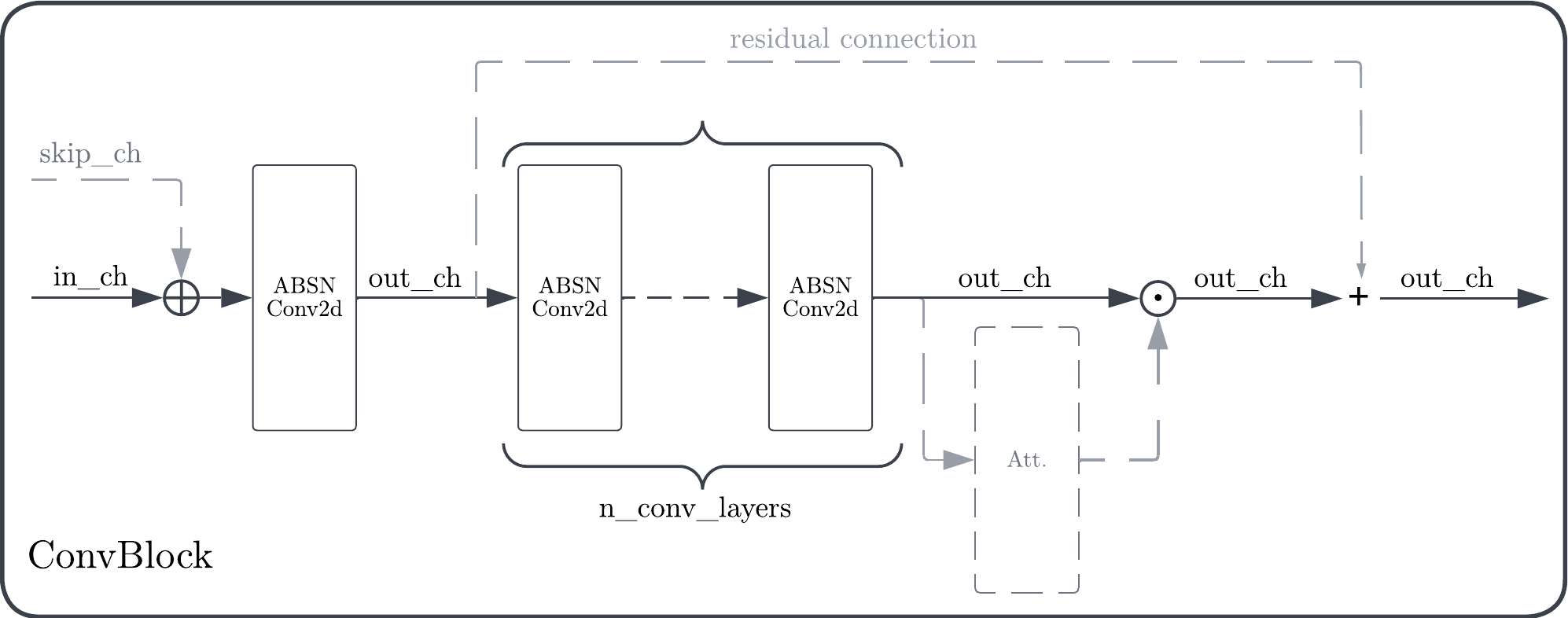}
    \caption{Convolutional block in the generator of Ui2i shown in Fig.~\ref{fig:generator}. Depending on the block's position within the network, optional components (marked with dashed gray lines) are enabled or omitted. The first convolutional layer adapts the number of channels and receives a skip connection. The output is then passed through a number of convolutional layers (equal to 2 in our experiments) without changing the number of channels. The resulting feature map is refined by the spatial-channel attention and residual connection. ABSN stands for approximate bidirectional spectral normalization.}
    \label{fig:convblock}
\end{figure*}

{\bf Generator.} The UNet-like architecture of the generators is composed of convolutional blocks, whose underlying design is illustrated in Fig.~\ref{fig:convblock}. The block consists of a number of convolutional layers with a kernel size of 3, along with the optional residual link and the optional channel-spatial attention module. Depending on the block's position within the network, the optional components of the block are enabled or omitted. Each encoder of the generators consists of three such blocks, followed by a downsampling layer implemented as a strided convolution (stride = 2), which both reduces the spatial resolution of the feature map and doubles the number of channels.  Each encoder feature is refined by a channel-spatial attention module, implemented as a sequence of the Efficient Symmetric Spatial and Channel Attention (ESCA) \cite{esca}, followed by the spatial attention \cite{cbam}. Additionally, each encoder block incorporates residual connections. In the UNet-like architecture, the encoders are followed by a shared bottleneck block, which omits both residual connections and attention mechanisms. The two decoders mirror the respective encoder's structure and incorporate skip connections at each level.  To mitigate ringing, aliasing, and checkerboard artifacts commonly introduced by standard upsampling layers, we adopt a $4\times4$ Lanczos2 kernel \cite{Turkowski90} for feature upscaling in the decoders, without reducing the number of channels in the input feature map. As a result, decoder features have twice as many channels as the corresponding encoder features (and their skip connections) at the same resolution level. Since decoder features are expected to be semantically rich, we omit the attention module in the decoder for efficiency, while retaining the residual connection.

{\bf Domain Discriminator.} Following PatchGAN architecture \cite{Isola17}, our Ui2i model uses a single domain discriminator with stacked images from two domains at the input. The initial layer is an $8\times8$ convolution with stride 2, increasing the number of feature channels to 128. This is followed by two additional $4\times4$ convolutions, also with stride 2, each doubling the number of features. Every contraction layer is followed by a leaky ReLU activation (with a negative slope of $0.2$). Subsequently, a $4\times4$ convolution with stride 1 further increases the number of features up to 1024. After another leaky ReLU activation, a final classification layer produces a feature map with either one or two channels, depending on whether or not an optional third class---corresponding to identity---is used in addition to the real and fake classes (class embeddings are explained in the Appendix).

{\bf Content discriminator} shares a similar, yet simplified, architecture compared to the domain discriminator. It uses a single contracting layer, with the number of channels fixed at 512 throughout, and a final classification layer that produces a feature map with one channel.

\section{Ui2i Training} 
\label{sec:learning}


During adversarial training of Ui2i, the domain discriminator is trained to classify real, fake and identity images, while the content discriminator learns to classify whether the bottleneck feature map originates from domain A or domain B. Simultaneously, the generators are optimized to fool the domain discriminator by making fake and identity images appear real, and to fool the content discriminator by making bottleneck features indistinguishable across the two domains.
Each training iteration randomly samples a pair of images from the two domains, ensuring they are content-unpaired. Following standard GAN training practices, we train Ui2i for $50,000$ iterations using the Adam optimizer with parameters $(\beta_1, \beta_2) = (0.5, 0.999)$ and a learning rate of $0.0002$. As in  \cite{CycleGAN}, we use image buffers that store the last 50 generated images from previous training iterations. We extend this approach by also maintaining buffers of bottleneck features from previous iterations. During training of the domain and content discriminators, we augment each current training sample with a randomly drawn sample from the corresponding buffer, propagating both together through the network. This strategy helps stabilize training by exposing the discriminators to a broader distribution of generated examples, reducing overfitting to the most recent outputs.

{\bf Loss Functions.} As in \cite{CycleGAN}, our training objective incorporates the following loss functions: (a) standard adversarial loss $\mathcal{L}_{\text{adv}}$ to supervise both the domain and content discriminators, as well as the two generators, within a min-max optimization framework;  (b) Cycle consistency loss $\mathcal{L}_\text{cyc}$; and (c) Optional identity loss $\mathcal{L}_\text{id}$ to enforce identity i2i translations when the generators are applied to images from their target domain (see Sec.~\ref{sec:content_preservation}). Additionally, to pull the bottleneck features of original and translated image pairs closer while pushing apart bottleneck representations of unrelated images, we use
the standard N-pair contrastive loss $\mathcal{L}_\text{cl}$ \cite{NPairLoss}. Thus, the total loss is defined as
\begin{equation}
    \mathcal{L} = \mathcal{L}_\text{adv} + \lambda_\text{cyc}\mathcal{L}_\text{cyc} + \lambda_\text{id}\mathcal{L}_\text{id} +\lambda_\text{cl}\mathcal{L}_\text{cl}.
    \label{eq:total_loss}
\end{equation}
Further details on each loss and the $\lambda$'s in Eq.~\eqref{eq:total_loss}  are provided in Appendix.

{\bf Differentiable Data Augmentation.} Following \cite{diffaug}, Ui2i training is regularized by applying augmentations to real, fake, and identity images when computing $\mathcal{L}_\text{adv}$. Specifically, to encourage scale-invariant feature learning, we randomly rescale each image using a scale factor sampled from the range $[0.75, 1.5]$. For scale factors below 1 (i.e., zooming out), reflection padding is applied to preserve the original image dimensions. As detailed in the next section, this augmentation significantly improves the performance of nuclear segmentation models on translated images.

\section{Ui2i for Domain Adaptation in Nuclear Segmentation}
\label{sec:segmentation}

\begin{figure}[ht]
    \centering
    \includegraphics[width=0.95\textwidth]{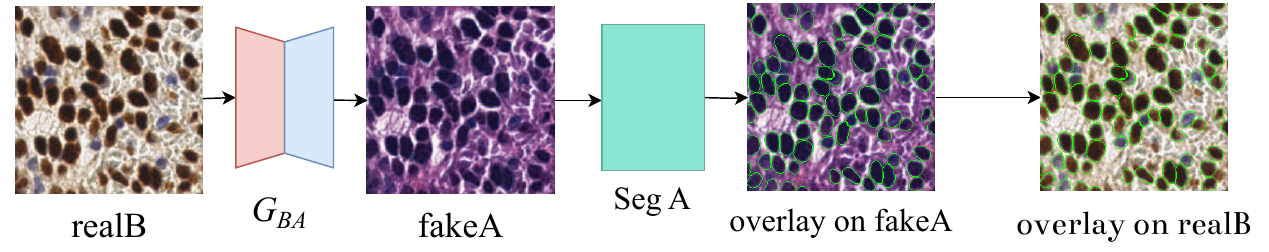}
    \caption{Our Ui2i model is trained to translate IHC images to the H\&E domain. Given an IHC input image (realB), the translated output (fakeA) is directly segmented by StarDist \cite{stardist}, pretrained on H\&E images (Seg A). Segmentation results (shown in green) are overlaid on the original input image.}
    \label{fig:pipeline}
\end{figure}

In biomedical applications, segmentation models pre-trained on one domain (e.g., IF, H\&E, or IHC) typically exhibit significant performance drop when applied to images from other domains. To avoid the need for retraining the segmentation model on the target domain, we instead use our Ui2i model to translate new images into the segmenter's original domain. This enables the segmentation model to be readily applied  to translated images without any additional training, as illustrated in Fig.~\ref{fig:pipeline}.

Tab.~\ref{tab:resheihc} presents nuclear segmentation results on IHC images using the evaluation metrics from \cite{stardist}: instance precision and recall based on intersection-over-union (IoU); segmentation quality estimated as the average IoU of instances with IoU > 0.5; and panoptic quality computed as the product of instance F1 score and segmentation quality. The evaluation is performed using the state-of-the-art (SOTA) model --- StarDist \cite{stardist, stardisthe}, pretrained in the H\&E domain on the MoNuSeg \cite{monuseg} dataset. Segmentation in the IHC domain without model re-training is made possible by translating input IHC images to the same H\&E style of the MoNuSeg dataset which was used for pretraining StarDist. For training Ui2i we use unpaired IHC and H\&E  datasets. The training IHC dataset consists of $7,918$ image patches with size $256\times256$ randomly sampled from the publicly available Lyon19 \cite{lyon19} and Endonuke \cite{endonuke} datasets. The training H\&E dataset consists of $2,368$  image patches with size $256\times256$ from the publicly available MoNuSeg \cite{monuseg} dataset. 
The test IHC dataset comprises 300 image patches from the Lyon19 and EndoNuke datasets, each annotated with expert ground-truth labels for nuclei. The loss weights are set as: $\lambda_\text{cyc}=10$, $\lambda_\text{id}=1$, and $\lambda_\text{cl} = 0.1$, and the setup with three-class domain discriminator is used. Training and testing on an NVIDIA GeForce RTX 4080 were performed under 5 hours and 1 minute, respectively. 

Tab.~\ref{tab:resheihc} compares alternative approaches for  IHC$\to$H\&E translation, including: 1) no translation---when StarDist and InstanSeg are applied directly to the input IHC images; 2) CycleGAN-based  IHC$\to$H\&E translation, and 3) i2i translations performed by the full Ui2i and its ablations. The Ui2i ablations evaluate the performance contributions of key components by removing one at a time: (a) Differentiable data augmentation (Ui2i w/o augment.); (b) Feature normalization replacing spectral normalization (Ui2i w/ feature norm.); and (c) Space-channel attention module (Ui2i w/o attention). Tab.~\ref{tab:resheihc} also includes an additional baseline---InstanSeg \cite{instanseg} pre-trained on a brightfield dataset that includes IHC images and is considered SOTA for the IHC domain.

As shown in Tab.~\ref{tab:resheihc}, Ui2i significantly improves instance precision and recall while maintaining segmentation quality compared to CycleGAN. Importantly, the translation also improves all H\&E-pretrained StarDist segmentation metrics over direct InstanSeg application to untranslated IHC images, despite InstanSeg being pretrained on the IHC domain. The Ui2i ablations further support the effectiveness of each proposed component.

\begin{table*}[h]
\centering
\caption{Results by StarDist \cite{stardist} pretrained in the H\&E domain and applied to translated H\&E images after IHC$\to$H\&E translation by: full Ui2i, Ui2i ablations, CycleGAN. The first row does not use translation. InstanSeg \cite{instanseg} is IHC-pretrained and directly applied to untranslated original IHC images.}
\label{tab:resheihc}
{\footnotesize
    \begin{tabular}{|c|c|c|c|c|c|c|}
  \hline
  \multicolumn{2}{|c|}{Setup} & Instance Precision & Instance Recall & Segm. Quality & Panoptic Quality \\
  \hline
 \multirow{2}{*}{No translation} & StarDist 
      & $0.92\pm0.13$ & $0.51\pm0.21$ & $0.78\pm0.08$ & $0.50\pm0.17$ \\
        \cline{2-6}
        & InstanSeg 
      & $0.76\pm0.16$ & $0.70\pm0.17$ & $0.75\pm0.08$ & $0.55\pm0.12$ \\
  \hline
     \hline
  \multicolumn{2}{|c|}{CycleGAN \cite{CycleGAN}}
      & $0.72\pm0.18$ & $0.76\pm0.16$ & $0.80\pm0.05$ & $0.59\pm0.14$ \\
  \hline
 \multicolumn{2}{|c|}{Ui2i w/o augment.}
      & $0.83\pm0.14$ & $0.72\mp0.14$ & $0.79\pm0.05$ & $0.60\pm0.11$ \\
  \hline
  \multicolumn{2}{|c|}{Ui2i w/ feature norm.}
      & $0.75\pm0.18$ & $0.74\pm0.16$ & $0.80\pm0.05$ & $0.60\pm0.14$ \\
  \hline
 \multicolumn{2}{|c|}{Ui2i w/o attention}
      & $0.83\pm0.14$ & $0.73\pm0.15$ & $0.80\pm0.05$ & $0.63\pm0.12$ \\
  \hline
  \multicolumn{2}{|c|}{full Ui2i}
      & $\boldsymbol{0.87\pm0.11}$ & $\boldsymbol{0.77\pm0.14}$ & $0.80\pm0.05$ & $\boldsymbol{0.65\pm0.10}$ \\
  \hline
\end{tabular}
}
\end{table*}

\section{Ui2i for Unmixing Single Channel IF Images}
\label{sec:demultiplexing}

In fluorescence microscopy, unmixing two markers imaged with the same fluorophore in a single channel effectively doubles the number of markers that can be visualized in a biological sample. To our knowledge, our system is the first capable of unmixing such signals using real-world single- and two-channel data. We present both qualitative and quantitative results to demonstrate its performance.

\begin{figure}
    \centering
    \includegraphics[width=0.9\linewidth]{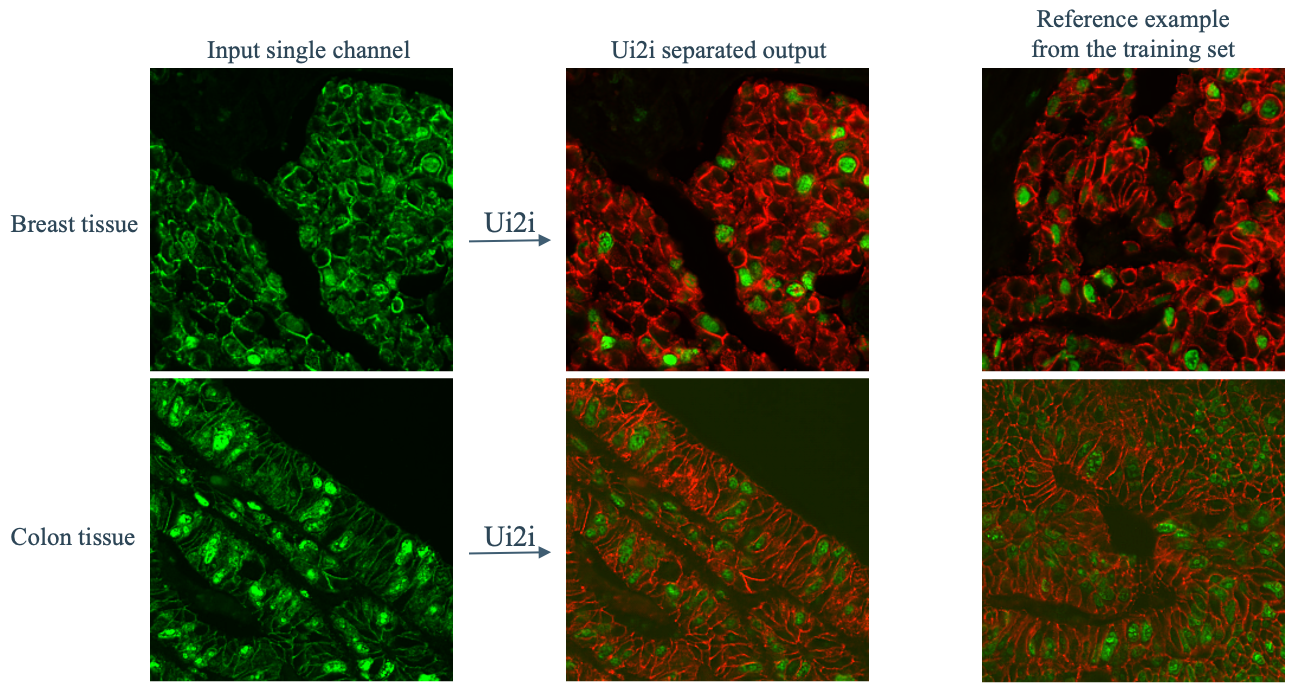}
    \caption{Breast and colon tissue unmixing results. Single-channel images, shown in the left column are separated into two channels (middle column). Example two-channel images from the training set are shown in the right column for reference.}
    \label{fig:UnmixingQualitative}
\end{figure}

We imaged human breast and colon tissue sections stained with Ki67 (nuclear marker) and E-cadherin (membranous marker), tagged with either distinct fluorophores (green and red, respectively) or the same fluorophore (green). This dataset reflects realistic staining and imaging conditions, but since ground truth (i.e., imaging the same sample with mixed and separated markers) is impossible, we provide qualitative assessments (Figure \ref{fig:UnmixingQualitative}). Breast tissue: 1,420 single-channel and 1,725 two-channel patches ($256\times256$ pixels). Colon tissue: 1,238 single-channel and 912 two-channel patches. Training: Completed in <5 hours; inference on all single-channel patches took <5 minutes. Loss weights: $\lambda_{\text{cyc}}=10$, $\lambda_{\text{cl}}=0.1$. Identity loss was excluded since it cannot be computed between single and multiple-channel images. 

For quantitative evaluation, we used the HT-T24 dataset from \cite{Ashesh2025.02.10.637323}, as it yielded the best results in their study. This three-channel dataset provides perfectly aligned, paired images of SOX2 (nuclear marker) and Grasp65 (Golgi marker), each tagged with distinct fluorophores. The imaging protocol involved sequential acquisition of individual markers followed by simultaneous excitation and imaging of both fluorophores. However, it should be noted that this approach produces multiplexed single-channel images that do not have the same characteristics of signal, noise and artifacts as the single-fluorophore multiplexing for which the unmixing solution is intended.

For training and evaluation, out of 7 available tissue sections, we randomly selected one for testing, 3 for unmixed and 3 for mixed domains, yielding a completely unpaired setup with $147$ mixed, $162$ unmixed and $57$ test images, all of size $1608\times1608$. During training, at each iteration, the current training sample is first resized to $804\times804$ pixels, and then a random crop of size $256\times256$ is taken before it is propagated through the networks. When the training is finished, we apply the resulting model to the test set and compare to ground truth to calculate metrics - peak signal to noise ratio (PSNR) and a modified structural similarity tailored for microscopy images, MicroMS-SSim \cite{microssim}. Since the mixed and unmixed domains have different number of channels, no identity loss is used. The loss weights are $\lambda_\text{cyc} = 5$, $\lambda_\text{cl}=0.1$. The hardware and the length of the training and inference remains the same as in the nuclear segmentation setup. We use only the full Ui2i setup, from previous section, without ablation studies. 

\begin{figure}
    \centering
    \includegraphics[width=1\linewidth]{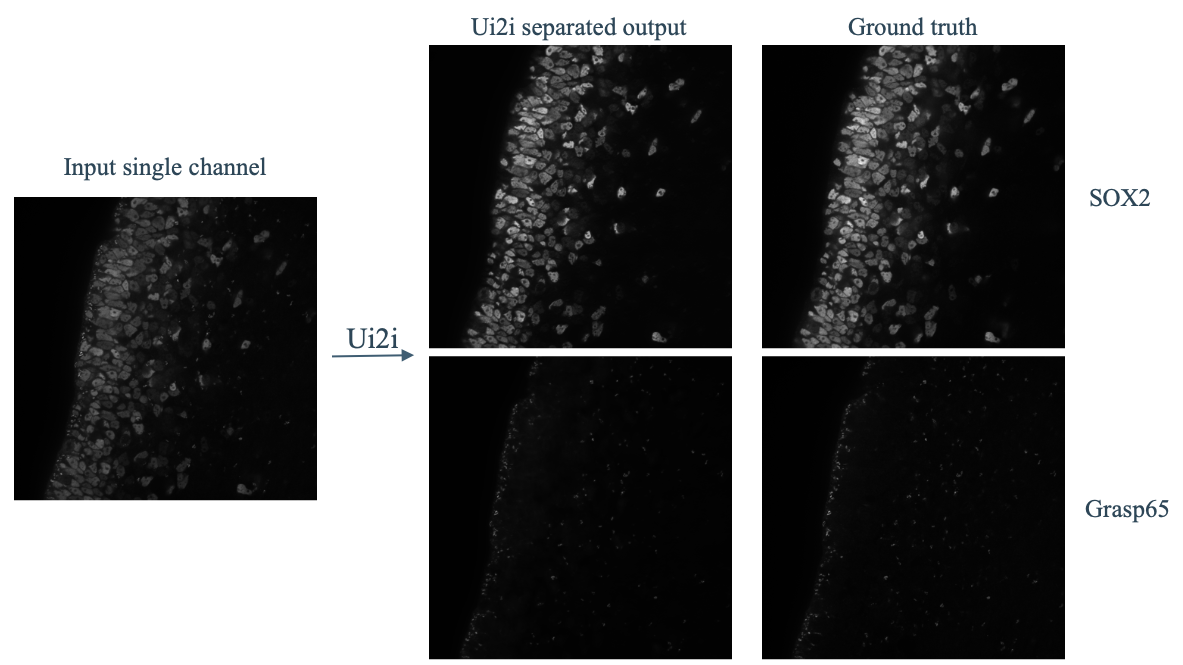}
    \caption{Unmixing results example on HT-T24. Single-channel input image is shown on the left and the unmixing result is shown in the middle (SOX2 is top, Grasp65 is bottom). Images on the right show the ground truth.}
    \label{fig:quantitativeUnmixing}
\end{figure}

Fig.~\ref{fig:quantitativeUnmixing} shows an example patch from the test set, the unmixing result and the corresponding ground truth. In Tab.~\ref{tab:resdemult}, we see very good results for unmixing the test section patches based on proposed metrics. For putting our results in context, MicroSplit \cite{Ashesh2025.02.10.637323}, which relies on paired and aligned training data, has similar performance on this dataset (MicroMS-SSIM: 0.978, 0.951 and PSNR: 40.3 and 32.8). Although, a direct comparison is challenging due to unclear train/test splits and out-of-focus sections they use. 

\begin{table*}[h]
\centering
\caption{Results for IF image demultiplexing on the test images. Peak signal to noise ratio (PSNR) and modified structural similarity (MicroSSIM) tailored to microscopy data for each channel.  }\label{tab:resdemult}
\label{tab:resdemultiplex}
{\footnotesize
    \begin{tabular}{|c|c|c|c|}
  \hline
  MicroMS-SSIM SOX2 & MicroMS-SSIM Grasp65 & PSNR SOX2 & PSNR Grasp65\\
   \hline
  $0.96\pm0.03$& $0.96\pm0.02$& $38\pm3$& $32\pm2$\\
  \hline
\end{tabular}
}
\end{table*}

\FloatBarrier
\section{Conclusion}
\label{sec:conclusion}
Domain adaptation for nuclear segmentation and unmixing in single-plex IF images are important applications of i2i translation requiring stricter structural preservation than typical tasks. To better preserve structural content, Ui2i extends prior work by: 1) replacing standard feature normalization with approximate bidirectional spectral normalization of network parameters; 2) incorporating the spatial-channel attention mechanisms and residual connections in the convolutional blocks of Ui2i; and 3) applying differentiable augmentations to training images for scale-invariant feature learning. Evaluation of nuclear segmentation on IHC images by applying H\&E-pretrained StarDist \cite{stardist, stardisthe} to translated H\&E images shows that our Ui2i-based IHC$\to$H\&E translation significantly improves instance precision and recall while maintaining segmentation quality compared to CycleGAN-based translation. Importantly, we also improve all H\&E-pretrained StarDist segmentation metrics over direct InstanSeg application to untranslated IHC images, despite InstanSeg being pretrained on the IHC domain. Results of signal unmixing in single-plex IF images further support the effectiveness of the proposed Ui2i. This is the first system that successfully separates superimposed IF markers, capable of training on real unpaired data. 

\section{Acknowledgements}
\label{sec:acknowledgements}
A provisional patent application related to this work has been filed.

\bibliographystyle{plain}
\bibliography{references}

\clearpage
\appendix

\begin{center}
    {\Large \bf Supplementary Material: Unpaired Image-to-Image Translation for Segmentation and Signal Unmixing}
\end{center}

\section{Domain and Content Discriminator Architectures}

\begin{figure*}[ht]
    \centering
    \includegraphics[width=.7\textwidth]{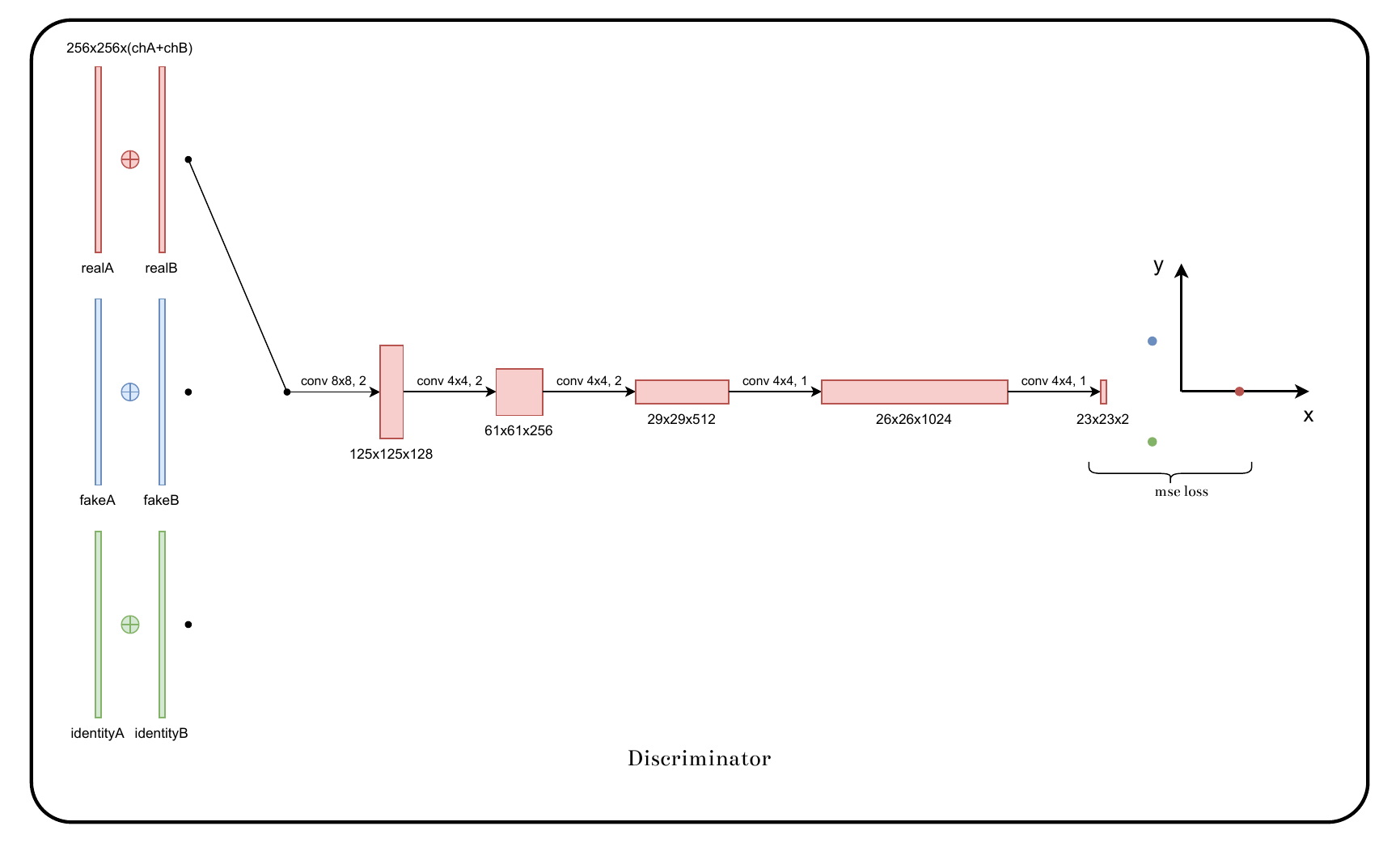}
    \caption{Domain discriminator. Notation: realA and realB denote images $x\in A$ and $y\in B$ sampled from their respective domains, fakeA and fakeB denote the translated images $x'=G_{BA}(y)$ and $y'=G_{AB}(x)$, respectively, and identityA and identityB denote the identity images $\hat{x}=G_{BA}(x)$ and $\hat{y}=G_{AB}(y)$.}
    \label{fig:discriminator}
\end{figure*}

\begin{figure*}[ht]
    \centering
    \includegraphics[width=.7\textwidth]{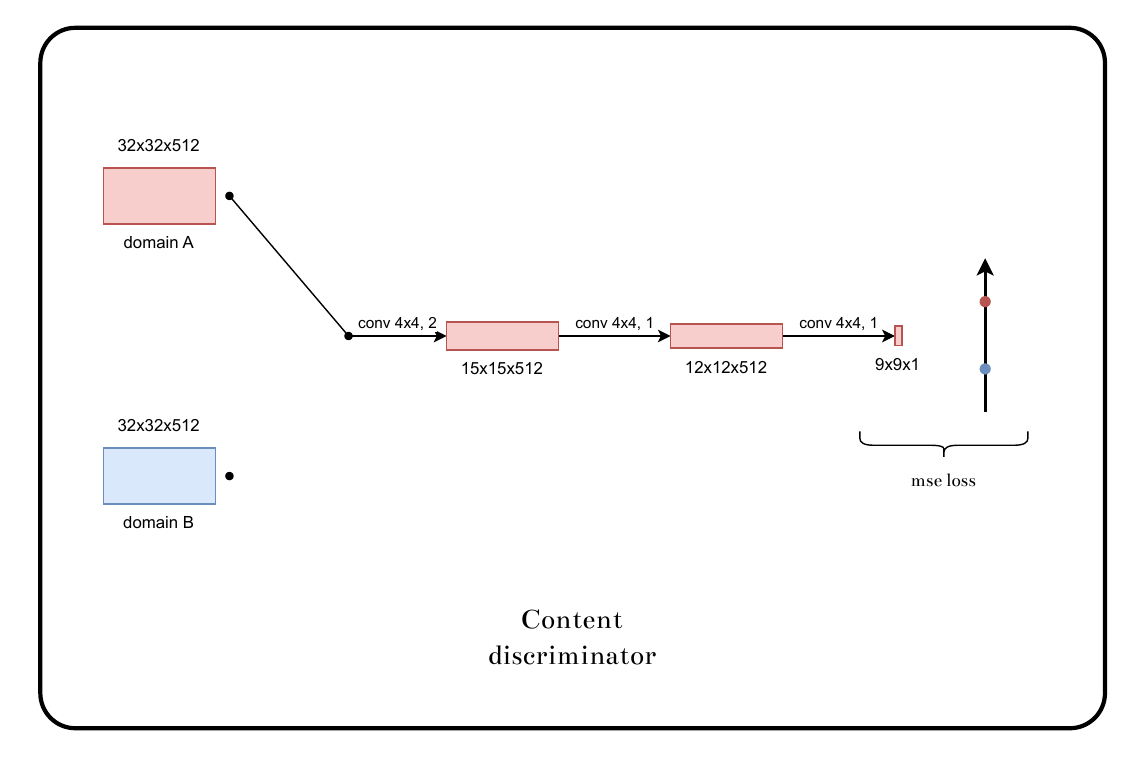}
    \caption{Content Discriminator}
    \label{fig:contentdisc}
\end{figure*}

Fig.~\ref{fig:discriminator} and Fig.~\ref{fig:contentdisc} show the detailed block diagrams of the domain discriminator and content discriminator, respectively. As shown in the figures, both discriminators are supervised by a mean squared loss, specified in the following section.


\section{Losses}\label{sec:losses}

This section uses the notation presented in the main paper to specify the expressions for all loss functions used for estimating the total loss, given by Eq.~(3), in the main paper.

{\bf Adversarial Loss} supervises the two generators, $G_{AB}$ and $G_{BA}$, and the two discriminators---domain discriminator $D_d$ and content discriminator $D_c$. Regular GANs \cite{GAN} treat the discriminator as a binary classifier and typically use the sigmoid cross-entropy loss for training. However, this approach is prone to vanishing gradients, which can hinder effective generator updates during training. Following LSGAN \cite{LSGAN}, we adopt the least squares loss function for training both the generators and discriminators. Minimizing the LSGAN loss amounts to minimizing the Pearson $\chi^2$ divergence.

The generators $G_{AB}$ and $G_{BA}$ translate images $x\in A$ and $y\in B$ as follows:
$$y'=G_{AB}(x), \quad x'=G_{BA}(y), \quad \hat{y}=G_{AB}(y), \quad \hat{x}=G_{BA}(x).$$ 
The resulting translations $x'$ and $y'$ are referred to as fake images, while $\hat{x}$ and $\hat{y}$ are  identity images.

As described in the main paper, following \cite{Isola17}, we employ a single domain discriminator $D_d$ for both domains A and B. To compute its loss, we stack an image pair $x \in A$ and $y \in B$ along the channel dimension---denoted as $[x|y]$---and feed this stacked input to $D_d$ to predict three classes: ``real'', ``fake'', and ``identity'' (where the ``identity'' prediction is omitted in the application of single-plex IF unmixing).  Let $D_{d}^r([x|y])$, $D_{d}^f([x|y])$, $D_{d}^i([x|y])$ denote the output scores for the respective three classes. To supervise these output scores with the least squares loss, we specify the (ground-truth) target scores 
$r$, $f$, and $i$ for the  ``real'', ``fake'', and ``identity'' classes, respectively, as coordinates of the three vertices of a 2D equilateral triangle: 
$$r = \left[\frac{\sqrt{3}}{2}, 0\right],\quad f = \left[-\frac{\sqrt{3}}{6}, \frac{1}{2}\right],\quad i = \left[-\frac{\sqrt{3}}{6}, -\frac{1}{2}\right].$$

On the other hand, the content discriminator $D_c$ predicts the originating domain of the input from its bottleneck representation, $z$. Let  $D_{c}^a(z)$ and $D_{c}^b(z)$ denote the output scores for domain A and B, respectively. To supervise these outputs, we specify the target scores as $a=1$ and $b=0$.

With the above specifications, the total adversarial loss  $\mathcal{L}_\text{adv}$---used in Eq.~(3) to train both the discriminators and generators---is defined as the sum of the following component losses:
\begin{align}
\mathcal{L}^{D_d}_\text{adv} &= \mathbb{E}_{\substack{x \in A\\y \in B}}\Big[\left|D_{d}^r([x|y]) - r\right|^2\nonumber + \abs{D_{d}^f([x'|y'])-f}^2 + \abs{D_{d}^i([x|\hat{x}]) - i}^2+\abs{D_{d}^i([y|\hat{y}]) - i}^2\Big],\\
\mathcal{L}^{D_c}_\text{adv} &= \mathbb{E}_{\substack{x \in A\\y \in B}}\Big[\left|D_{c}^a(z(x)) - a\right|^2\nonumber + \abs{D_{c}^b(z(y))-b}^2\Big],\\
     \mathcal{L}^{G}_\text{adv} &= \mathbb{E}_{\substack{x \in A\\y \in B}}\left[\left|D_{d}^f([x'|y']) - r\right|^2 + \abs{D_{d}^i([x|\hat{x}]) - r}^2+\abs{D_{d}^i([y|\hat{y}]) - r}^2 \right.\\
     &\quad \quad\quad\;+ \left.\abs{D_{c}^a(z(x)) - \frac{a+b}{2}}^2+ \abs{D_{c}^b(z(y)) - \frac{a+b}{2}}^2\right].
\end{align}

{\bf Cycle Consistency Loss.}

\[ \mathcal{L}_\text{cyc} = \mathbb{E}_{x\in A} \big[\abs{x - G_{BA}(G_{AB}(x))}\big] + \mathbb{E}_{y\in B}\big[\abs{y - G_{AB}(G_{BA}(y))}\big]\]

{\bf Identity Loss.} 
\begin{equation}
    \mathcal{L}_\text{id} = \mathbb{E}_{x\in A}\big[\abs{x - G_{BA}(x)}\big] + \mathbb{E}_{y\in B}\big[\abs{y - G_{AB}(y)}\big].
\end{equation}

When the two domains are close—for example, when bright and dark regions correspond across input and output, or when only specific objects require modification while the overall scene remains unchanged—identity loss is particularly effective for content preservation. Conversely, when the domains are not close, identity loss may hinder content preservation.

{\bf Cross-domain Contrastive Loss.} To enhance representation learning, the bottleneck block of the UI2I model is additionally supervised with a cross-domain contrastive loss, as in \cite{mixdomaincontrastive, patchwisecontrastive}. To preserve the representational capacity of the bottleneck feature map, $z$, the contrastive loss is not applied directly to $z$. Instead, $z$ is passed through a multilayer perceptron (MLP) consisting of a linear layer, a ReLU activation, a second linear layer, and $\ell_2$ normalization, resulting in $\tilde{z}$. Departing from the standard practice of using a globally average-pooled $\tilde{z}$ to compute the loss, we instead average the per-pixel loss components over the entire feature map $\tilde{z}$. Thus, for each pixel in $\tilde{z}$, we first compute the standard N-pair loss over the anchor image $u$, positive image $u^+$, and a set of negative images $\{u_i^-\}$ as:
\begin{equation}\label{eq:contrpix}
    \mathcal{L}_{\text{cl}}(u, u^+, \{u_i^-\}) = \mathbb{E}_{u}\left[ -\log\frac{e^{\tilde{z}^\top(u)\cdot \tilde{z}(u^+)/\tau}}{e^{\tilde{z}^\top(u)\cdot \tilde{z}(u^+)/\tau} + \sum_{i}e^{\tilde{z}^\top(u)\cdot \tilde{z}(u_i^-)/\tau}}\right],
\end{equation}
for each of the following four types of anchors and their corresponding positives and negatives:
\begin{subequations}
\begin{align}
    &u = x \in A,\quad &&u^+ = G_{AB}(x),\quad &&\{u_i^-\} = \{x_1 \in A, y \in B, y_1 \in B, G_{BA}(y)\} \\
    &u = y \in B,\quad &&u^+ = G_{BA}(y),\quad &&\{u_i^-\} = \{y_1 \in B, x \in A, x_1 \in A, G_{AB}(x)\} \\
    &u = G_{BA}(y),\ y \in B,\quad &&u^+ = y,\quad &&\{u_i^-\} = \{y_1 \in B, x \in A, x_1 \in A, G_{AB}(x)\} \\
    &u = G_{BA}(x),\ x \in A,\quad &&u^+ = x,\quad &&\{u_i^-\} = \{x_1 \in A, y \in B, y_1 \in B, G_{BA}(y)\}
\end{align}
\end{subequations}
where $x_1\in A$ and $y_1\in B$ are additional images selected from domains A and B, respectively, such that $x\ne x_1$ and $y\ne y_1$. Then, we average $\mathcal{L}_{\text{cl}}(u, u^+, \{u_i^-\})$ given by Eq.~\eqref{eq:contrpix} across all pixels in $\tilde{z}$ to compute the cross-domain contrastive loss. 

It is worth noting that pixelwise contrastive loss for unpaired image-to-image translation, introduced in \cite{patchwisecontrastive}, selects anchor and positive samples as pixel-aligned features from a ResNet-based encoder, while negative samples are drawn from other pixels within the same feature map. In our case, however, the use of UNet-based generators results in a larger receptive field, leading to greater overlap between neighboring feature vectors and reducing the effectiveness of intra-image negative mining. Instead, inspired by \cite{mixdomaincontrastive}, we sample negatives from the deep features of different images. For each anchor-positive pair, all pixels from these external images are treated as negative examples.


\section{Parameter Initialization}

Weights of all layers are initialized by sampling the uniform distribution $\mathcal{U}[-2\sqrt{3}/10, 2\sqrt{3}/10]$, while biases are initialized by sampling the Xavier uniform distribution. Sampling all initial weights from the same distribution is suitable, because the spectral normalization introduced in the generators plays a role of learning rate equalizer \cite{progan}. To accelerate the initial phase of learning, bias of the final layer is initialized such that its activation---hyperbolic tangent---equals the median of the pixel value of the target domain of the generator.

\section{Spatial-Channel Attention Module}\label{sec:attention}
 
 For the spatial-channel attention modules used in the generators, we resort to ESCA \cite{esca}. ESCA uses the following sequence of attention processing: channel attention$\to$height attention$\to$width attention$\to$channel attention. We extend ESCA by adding one more layer that is very similar to the spatial attention from CBAM \cite{cbam}. Instead of concatenating both the average and max pooling of responses along the channel dimension, we concatenate the average and root-mean-square pooling of responses to facilitate gradient flow and stable training.

\section{Impact of Feature Normalization Layers: Toy Example}

This section motivates our choice of spectral normalization over conventional feature normalization through a toy example. We show that instance normalization, as one way of performing feature normalization, can cause the response of a localized object to vary significantly with its surrounding context---an effect reflecting the variability commonly observed in biomedical images. Such context-dependent responses, stemming from feature normalization's reliance on global statistics, are undesirable for applications requiring strict structural fidelity, like ours. To overcome this, we replace feature normalization with spectral normalization, which ensures consistent responses regardless of context.

\begin{figure}[ht]
    \centering
    \includegraphics[width=0.95\linewidth]{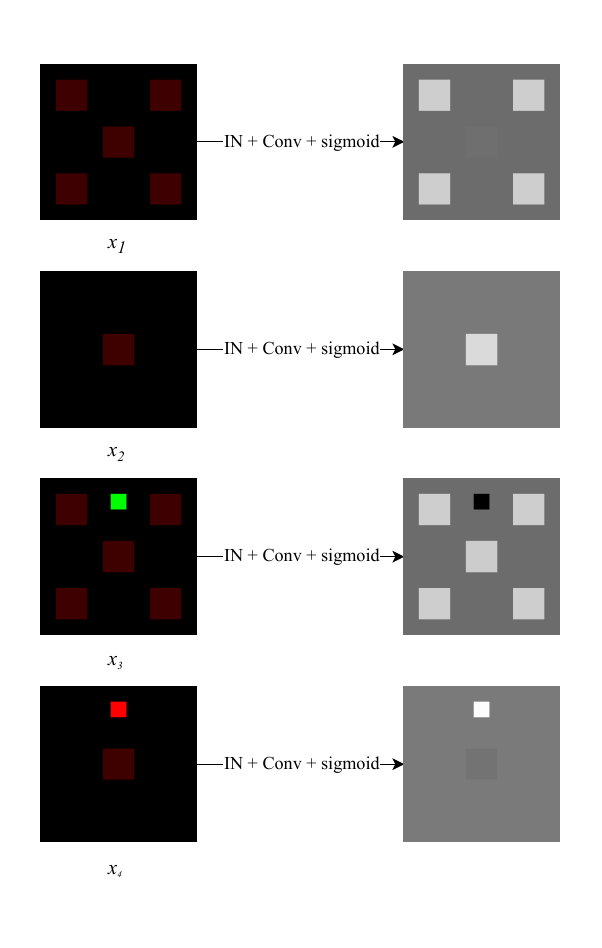}
    \caption{Toy example illustrating how instance normalization (as a special case of feature normalization) causes the response of localized objects to vary with image context due to its reliance on global feature statistics. The four two-channel inputs ($x_1$, $x_2$, $x_3$, $x_4$) are visualized as RGB images with the blue channel set to zero. The corresponding single-channel responses (after instance normalization, $1{\times}1$ convolution, and sigmoid activation) are shown in grayscale. Notably, the central square---whose second channel value is small relative to the first---exhibits markedly different responses depending on the surrounding content.}
    \label{fig:blobstoy}
\end{figure}

Fig.~\ref{fig:blobstoy} shows four two-channeled feature maps, $x_1, x_2, x_3, x_4$, each with a spatial size of $100\times100$, and zero-valued everywhere in both channels except at certain regions. 
Specifically:
\begin{enumerate}
    \item $x_1$ contains four $20 \times 20$ squares near the corners, where the first channel has a value of $0.25$ and the second channel is $0$. Additionally, $x_1$ has a central square of the same size, where the first channel remains $0.25$, but the second channel is set to $0.00875$. 
    \item $x_2$ has only the central square, same as in $x_1$.
    \item $x_3$ has the same squares as in $x_1$ and adds another sixth square of size $10\times10$ where the first channel has intensity $0$ and the second channel has intensity $1$.
    \item $x_4$ adds an additional square to $x_2$ of size $10\times10$ where the first channel has intensity $1$ and the second channel has intensity $0$. 
\end{enumerate}

Consider applying an instance normalization layer to each of the four feature maps, followed by a $1 \times 1$ convolution that reduces the number of channels to 1. The convolution uses a weight vector $\vec{w} = \frac{1}{\sqrt{2}}(1, -1)$, and is followed by a sigmoid activation function.

As shown in Fig.~\ref{fig:blobstoy}, instance normalization causes the response of localized objects to vary significantly depending on the surrounding context, due to its reliance on global feature statistics. The presence of high-intensity regions elsewhere in the image alters how a localized object is interpreted. In contrast, spectral normalization is parameter-based and does not depend on global image context, resulting in consistent responses for localized regions, regardless of surrounding image content.

\section{Domain Adaptation for Nuclear Segmentation Experiments}

We evaluated the performance of the Ui2i system on a domain adaptation task for nuclear segmentation in IHC images. A total of 7,918 image patches ($256\times256$ pixels) were randomly sampled from the Lyon19 \cite{lyon19} and EndoNuke \cite{endonuke} datasets to form the IHC training set. An independent test set of 300 patches from the same datasets was randomly selected and manually annotated by an expert.

Two segmentation models were used for comparison: StarDist \cite{stardist, stardisthe}, trained exclusively on H\&E-stained images from the MoNuSeg dataset \cite{monuseg} and InstanSeg \cite{instanseg}, a state-of-the-art model trained on a mixture of H\&E and IHC images from five datasets (TNBC-2018\cite{TNBC_dataset}, LyNSeC \cite{LyNSeC}, NuInsSeg \cite{NuInsSeg}, IHC -TMA \cite{IHC_TMA} and CoNSeP \cite{CoNSep}).
We evaluated four approaches:
\begin{enumerate}
    \item StarDist applied directly to test IHC images
    \item CycleGAN-based translation from IHC to H\&E, followed by StarDist segmentation
    \item Ui2i-based translation from IHC to H\&E, followed by StarDist segmentation
    \item InstanSeg applied directly to test IHC images
\end{enumerate}
Both translation models - CycleGAN and Ui2i - were trained to map the IHC training set into the H\&E domain of the MoNuSeg dataset, the exact domain StarDist was trained on. 

\begin{figure*}[htb]
    \centering
    
    \begin{subfigure}[t]{\textwidth}
        \begin{minipage}[c]{0.18\textwidth}
            \subcaption{Examples from test dataset.}
            \label{fig:test-examples1}
        \end{minipage}\hfill
        \begin{minipage}[c]{0.80\textwidth}
            \centering
            \includegraphics[width=0.22\textwidth]{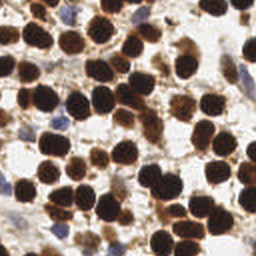}%
            \hfill
            \includegraphics[width=0.22\textwidth]{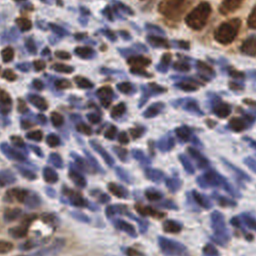}%
            \hfill
            \includegraphics[width=0.22\textwidth]{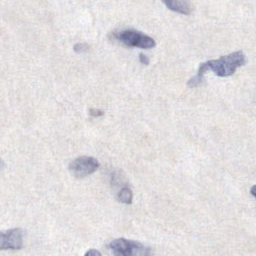}%
            \hfill
            \includegraphics[width=0.22\textwidth]{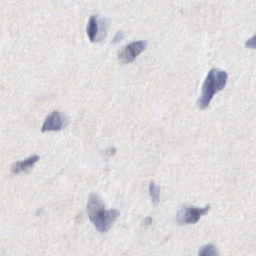}
        \end{minipage}
    \end{subfigure}
    
    \vspace{10pt}

    \begin{subfigure}[t]{\textwidth}
        \begin{minipage}[c]{0.18\textwidth}
            \subcaption{Ground truth segmentation.}
            \label{fig:test-examples1}
        \end{minipage}\hfill
        \begin{minipage}[c]{0.80\textwidth}
            \centering
            \includegraphics[width=0.22\textwidth]{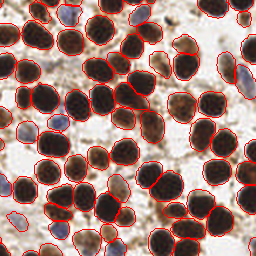}%
            \hfill
            \includegraphics[width=0.22\textwidth]{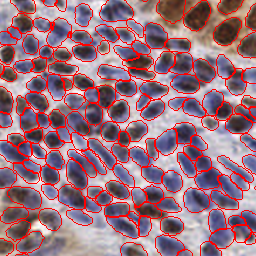}%
            \hfill
            \includegraphics[width=0.22\textwidth]{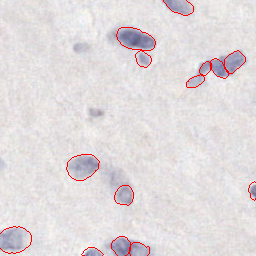}%
            \hfill
            \includegraphics[width=0.22\textwidth]{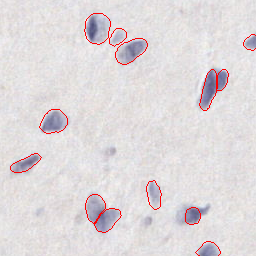}
        \end{minipage}
    \end{subfigure}
    
    \vspace{10pt}

    \begin{subfigure}[t]{\textwidth}
        \begin{minipage}[c]{0.18\textwidth}
            \subcaption{No translation + StarDist.}
            \label{fig:test-examples1}
        \end{minipage}\hfill
        \begin{minipage}[c]{0.80\textwidth}
            \centering
            \includegraphics[width=0.22\textwidth]{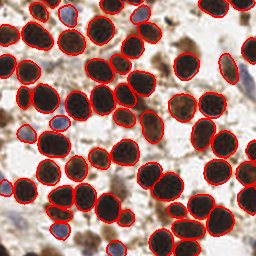}%
            \hfill
            \includegraphics[width=0.22\textwidth]{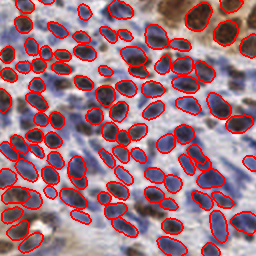}%
            \hfill
            \includegraphics[width=0.22\textwidth]{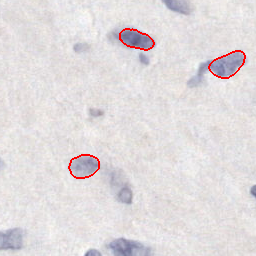}%
            \hfill
            \includegraphics[width=0.22\textwidth]{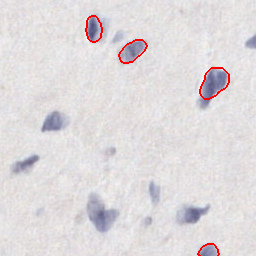}
        \end{minipage}
    \end{subfigure}
    
    \vspace{10pt}

    \begin{subfigure}[t]{\textwidth}
        \begin{minipage}[c]{0.18\textwidth}
            \subcaption{No translation + InstanSeg.}
            \label{fig:test-examples1}
        \end{minipage}\hfill
        \begin{minipage}[c]{0.80\textwidth}
            \centering
            \includegraphics[width=0.22\textwidth]{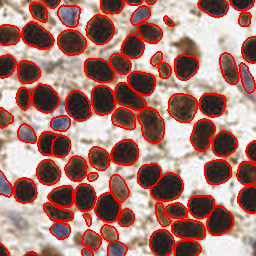}%
            \hfill
            \includegraphics[width=0.22\textwidth]{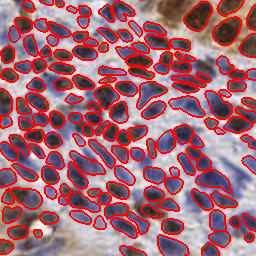}%
            \hfill
            \includegraphics[width=0.22\textwidth]{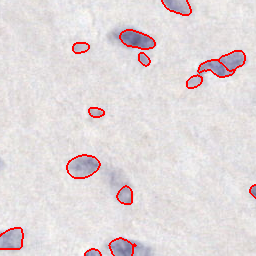}%
            \hfill
            \includegraphics[width=0.22\textwidth]{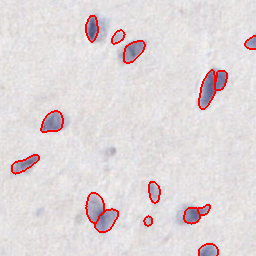}
        \end{minipage}
    \end{subfigure}
    
    \vspace{10pt}
    
    \begin{subfigure}[t]{\textwidth}
        \begin{minipage}[c]{0.18\textwidth}
            \subcaption{Cycle Gan translation.}
            \label{fig:test-examples2}
        \end{minipage}\hfill
        \begin{minipage}[c]{0.80\textwidth}
            \centering
            \includegraphics[width=0.22\textwidth]{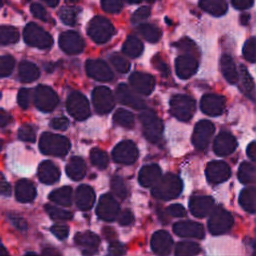}%
            \hfill
            \includegraphics[width=0.22\textwidth]{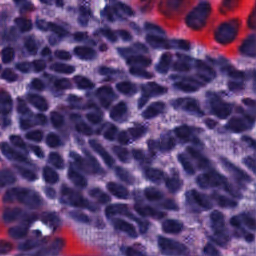}%
            \hfill
            \includegraphics[width=0.22\textwidth]{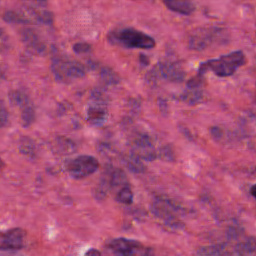}%
            \hfill
            \includegraphics[width=0.22\textwidth]{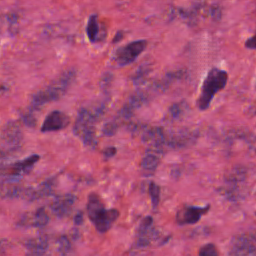}
        \end{minipage}
    \end{subfigure}
    
    \vspace{10pt}

    \begin{subfigure}[t]{\textwidth}
        \begin{minipage}[c]{0.18\textwidth}
            \subcaption{Cycle Gan translation + StarDist.}
            \label{fig:test-examples-CycleGantranslation}
        \end{minipage}\hfill
        \begin{minipage}[c]{0.80\textwidth}
            \centering
            \includegraphics[width=0.22\textwidth]{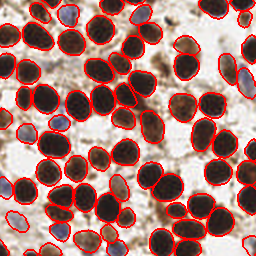}%
            \hfill
            \includegraphics[width=0.22\textwidth]{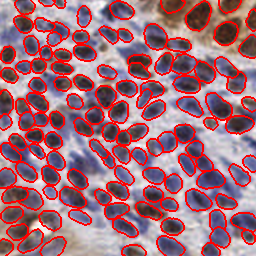}%
            \hfill
            \includegraphics[width=0.22\textwidth]{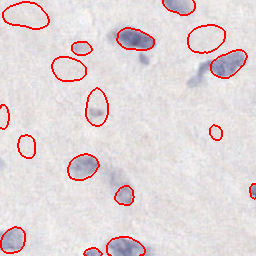}%
            \hfill
            \includegraphics[width=0.22\textwidth]{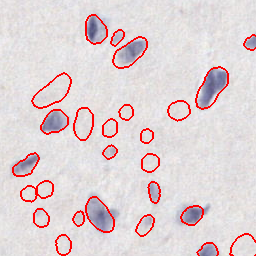}
        \end{minipage}
    \end{subfigure}
    
    \vspace{10pt}

    \begin{subfigure}[t]{\textwidth}
        \begin{minipage}[c]{0.18\textwidth}
            \subcaption{UI2I translation.}
            \label{fig:setup22}
        \end{minipage}\hfill
        \begin{minipage}[c]{0.80\textwidth}
            \centering
            \includegraphics[width=0.22\textwidth]{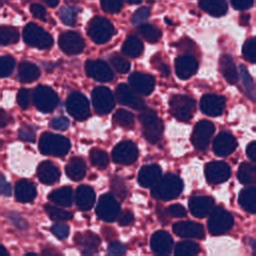}%
            \hfill
            \includegraphics[width=0.22\textwidth]{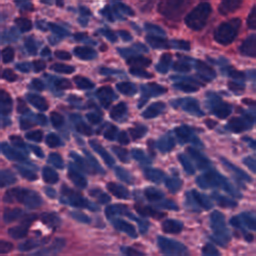}%
            \hfill
            \includegraphics[width=0.22\textwidth]{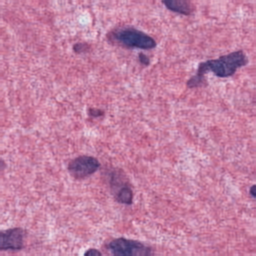}%
            \hfill
            \includegraphics[width=0.22\textwidth]{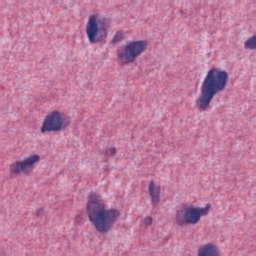}
        \end{minipage}
    \end{subfigure}
    
    \vspace{10pt}

    \begin{subfigure}[t]{\textwidth}
        \begin{minipage}[c]{0.18\textwidth}
            \subcaption{UI2I translation + StarDist.}
            \label{fig:setup2}
        \end{minipage}\hfill
        \begin{minipage}[c]{0.80\textwidth}
            \centering
            \includegraphics[width=0.22\textwidth]{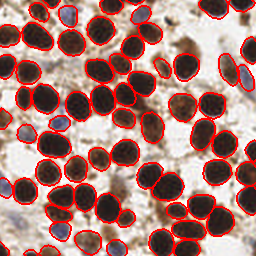}%
            \hfill
            \includegraphics[width=0.22\textwidth]{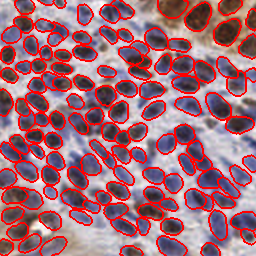}%
            \hfill
            \includegraphics[width=0.22\textwidth]{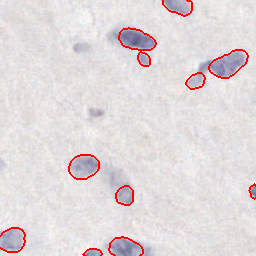}%
            \hfill
            \includegraphics[width=0.22\textwidth]{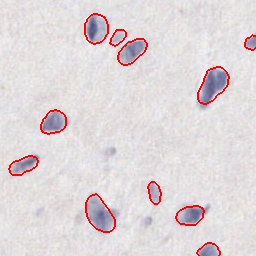}
        \end{minipage}
    \end{subfigure}

    \caption{ Domain adaptation for nuclear segmentation experiment: The top row shows example patches from the original IHC test images, illustrating the differences in color, shape, size and density of nuclei seen in the test dataset. CycleGAN struggles with hallucinations that are particularly prominent in low density patches}
    \label{fig:image-grid}
\end{figure*}

Figure \ref{fig:image-grid} presents results from the four approaches on representative image patches selected to illustrate the diversity of appearance, including variation in shape, size, density, and staining patterns. CycleGAN exhibits visible artifacts and hallucinated structures, particularly in regions with low nuclear density.

It is important to note that translating IHC images into the domain of a specialized model - specifically, the H\&E domain that StarDist was trained on - yields better segmentation performance than the state-of-the-art generalist model (InstanSeg) that was trained on a broader mix of H\&E and IHC stains. This highlights the utility of precise domain alignment over generalization across datasets.

\section{Ui2i for Unmixing Single Channel IF Images Experiments}

\begin{figure}[htb!]
    \centering
    \includegraphics[width=1\linewidth]{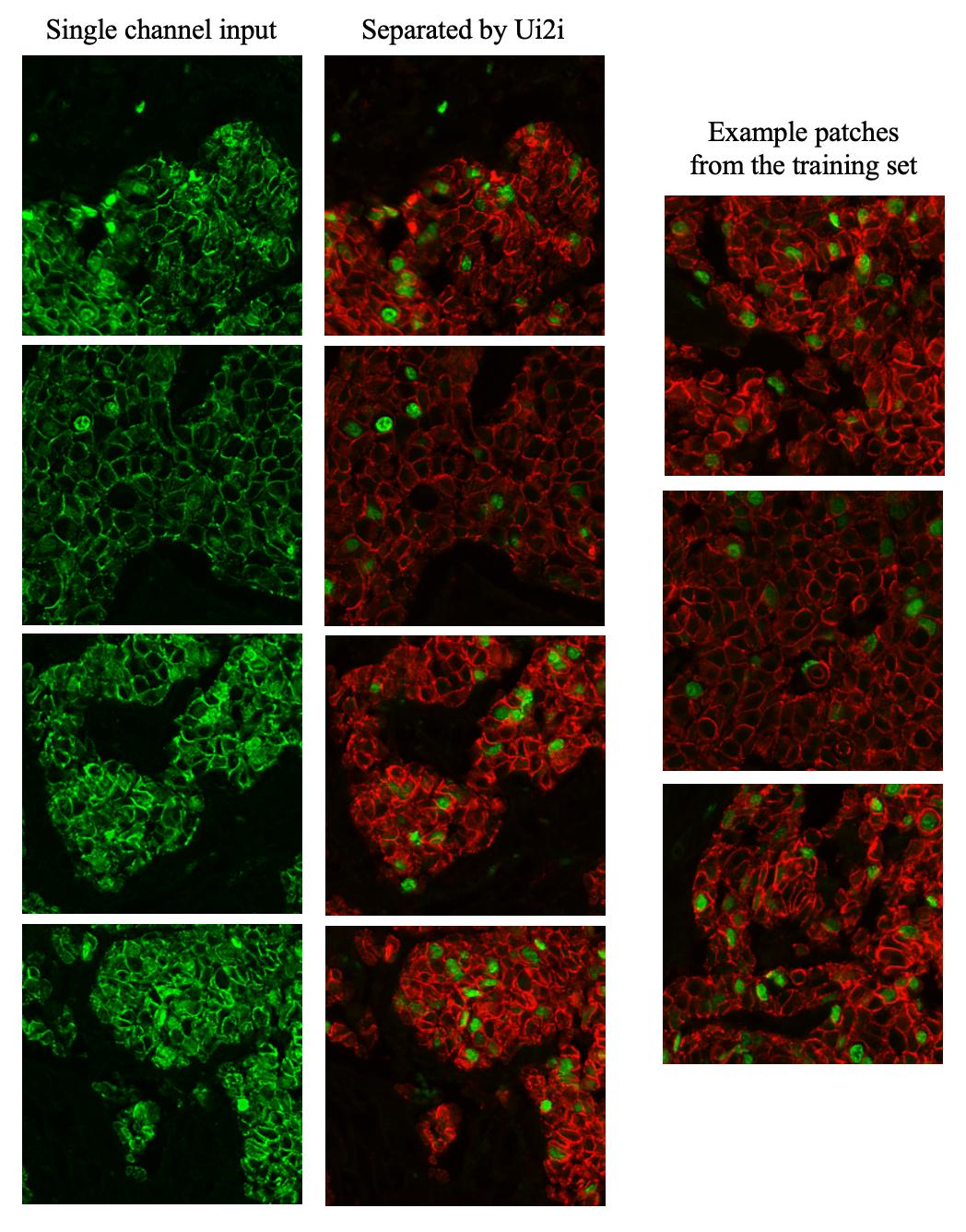}
    \caption{Example patches from the single-channel IF separation in breast tissue. Left column shows the input single-channel multiplexed images. The middle column shows the channels separated by our algorithm. Sample patches from the training set are shown on the right for reference.}
    \label{fig:breast-example}
\end{figure}

\begin{figure}[htb!]
    \centering
            \includegraphics[width=1\linewidth]{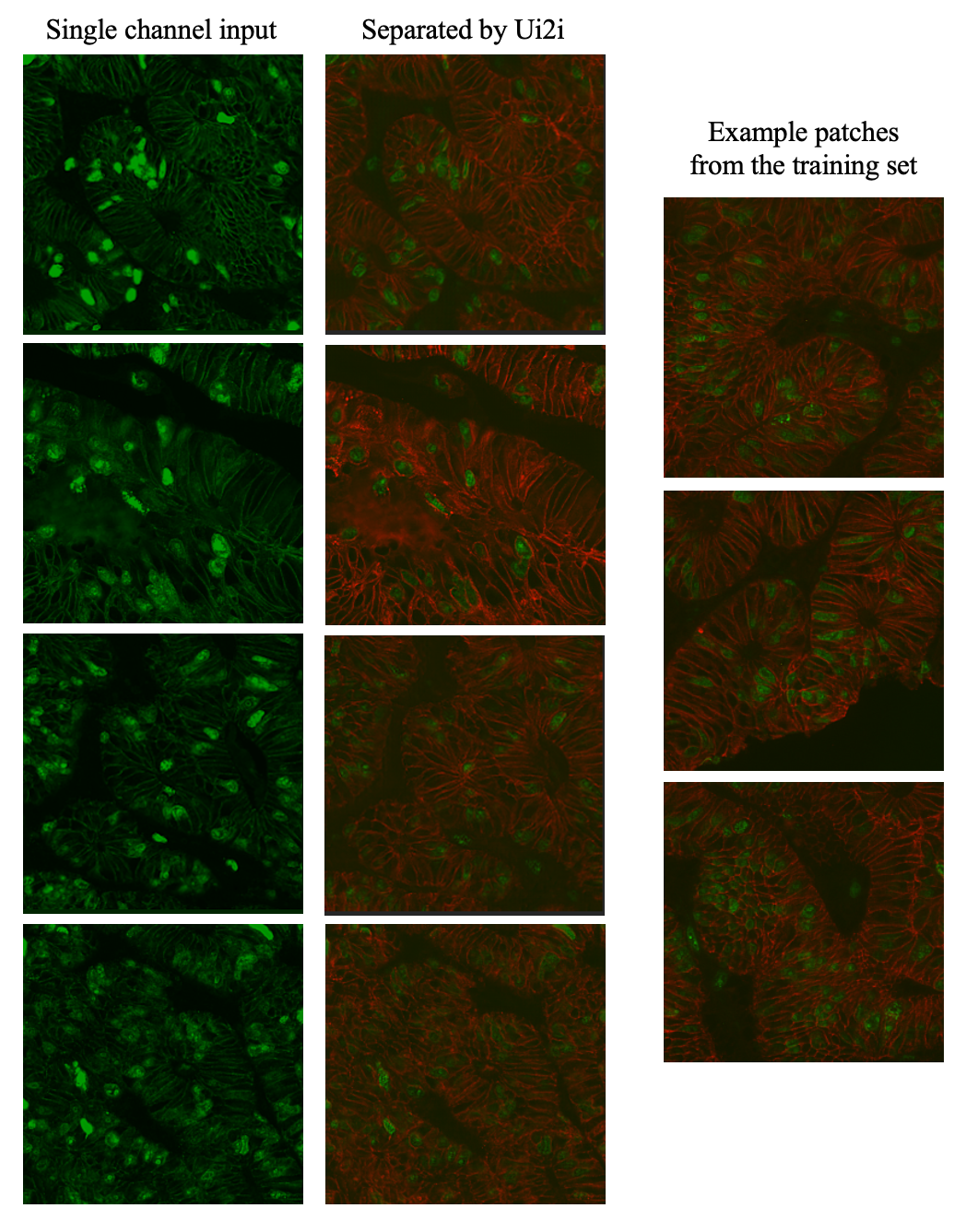}
                  \caption{Example patches from the single-channel IF separation in colon tissue. Left column shows the input single-channel multiplexed images. The middle column shows the channels separated by our algorithm. Sample patches from the training set are shown on the right for reference.}
    \label{fig:colon-example}
        
\end{figure}

We include additional example patches from both experiments to further illustrate our results. In the first experiment, the single-channel datasets for human breast and colon tissue were acquired using the same protocol as the intended application of our system: the same fluorophore was used to label both Ki67 and E-cadherin, and a single-channel image was acquired (alongside a DAPI channel, which is standard but omitted here for clarity in visualizing separation quality). For training, we randomly sampled patches of $256\times256$ pixels. For breast tissue we had 1,420 single-channel and 1,725 two-channel patches and for colon tissue 1,238 single-channel and 912 two-channel patches. Since ground truth separation is not possible to obtain for these images, we instead show the output of our model alongside example real two-channel images from the training set for reference. Example patches from breast and colon tissue are shown in Figures~\ref{fig:breast-example}~and~\ref{fig:colon-example}, respectively.

\begin{figure}[htb!]
    \centering
    \includegraphics[width=1\linewidth]{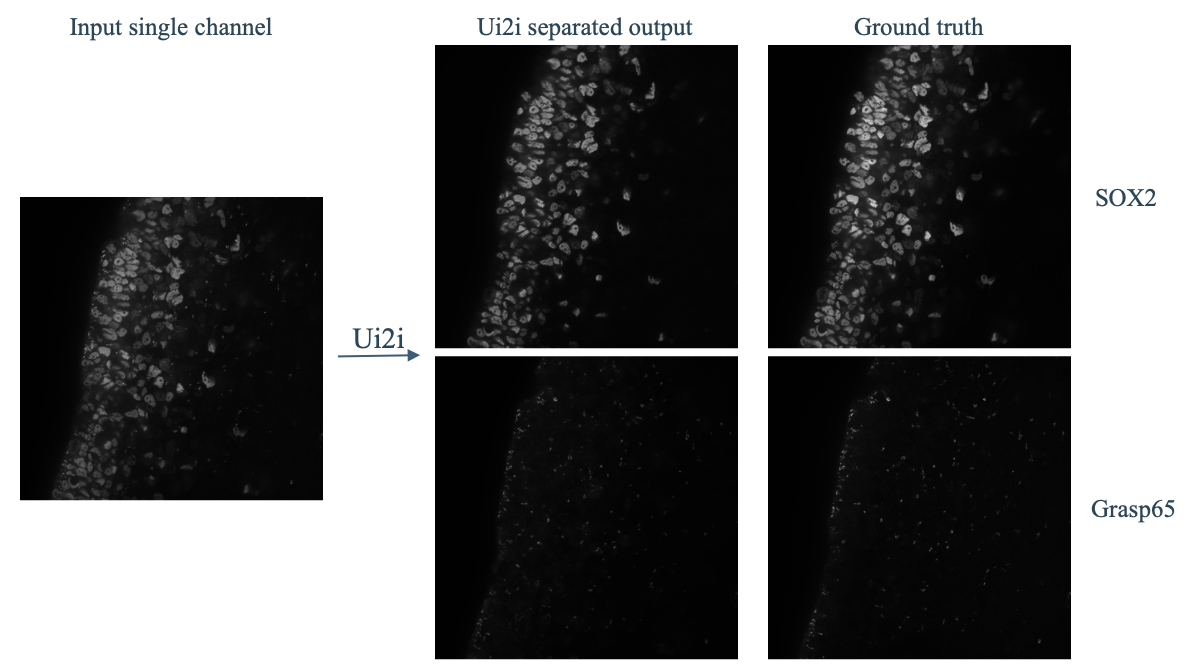}
\hfill
    \includegraphics[width=1\linewidth]{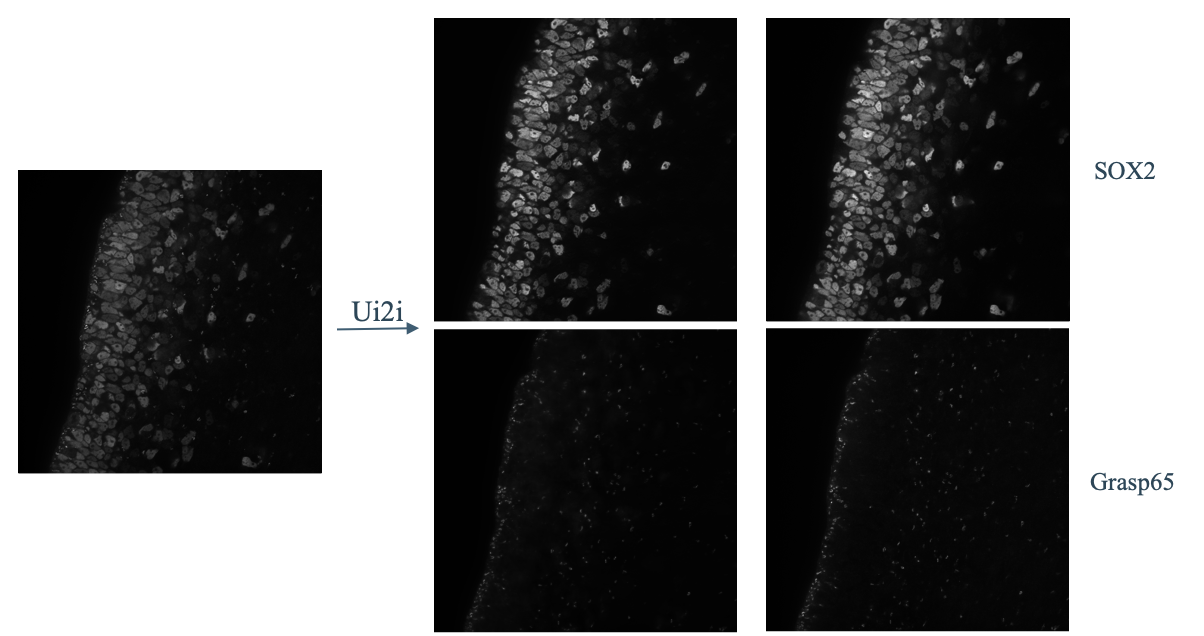}
\hfill
        \includegraphics[width=1\linewidth]{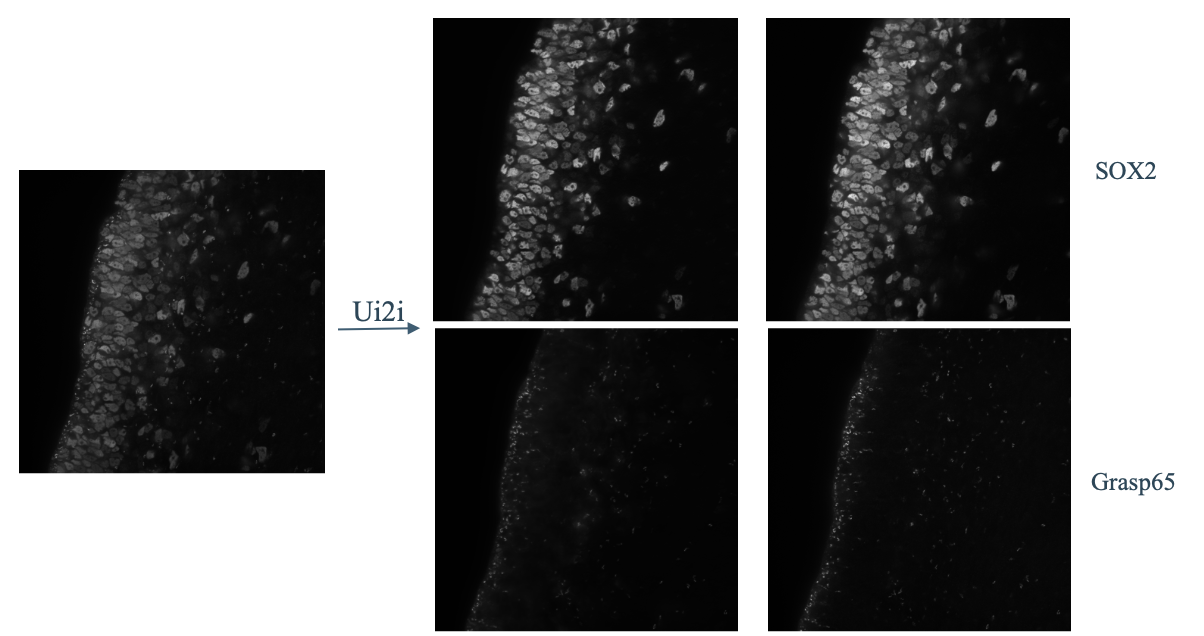}
   
    \caption{Additional example patches for IF marker separation for the dataset with available ground truth.}
    \label{fig:paired examples}
\end{figure}

Figure\ref{fig:paired examples} presents additional examples from the second experiment, using the HT-T24 data set from \cite{microsplitcode}. This dataset is a three channel confocal microscopy imaging of ferret brain sections. It is a 3-channel 2D dataset. First two channels contain the SOX2 (transcription factor used to label stem cells nuclei) and Golgi marker Grasp65. The dataset consists of seven sections, one was randomly chosen for testing (Test1\_Slice4\_b), three for single-channel (Test1\_Slice3\_b, Test1\_Slice1, Test1\_Slice4\_a) and three for two-channel (Test1\_Slice3\_a, Test1\_Slice2\_b, Test1\_Slice2\_a) training datasets. Original images of size 1608x1608 were resized to 804x804 and then random crops of size $256\times256$ with random horizontal/vertical flips were taken during training. This dataset includes perfectly aligned three-channel images, acquired through a sequential imaging protocol: each marker was imaged separately, followed by simultaneous excitation and acquisition of both fluorophores in a single multiplexed channel. This is a reasonable and practical strategy when working with paired image-to-image translation systems, as it provides precisely aligned input-output examples for supervised learning. However, it’s important to note that the resulting multiplexed single-channel images do not faithfully reproduce the signal characteristics, noise patterns, or artifact profiles of true single-fluorophore multiplexing, the specific use case our unmixing system is designed to handle. These images are shown in grayscale, as the Grasp65 marker is difficult to distinguish in composite two-color views.

\section{Failure Example: Non-Overlapping Content Spaces}

\begin{figure}[htb!]
    \centering
        \begin{minipage}[c]{0.99\textwidth}
            \centering
            \includegraphics[width=0.33\textwidth]{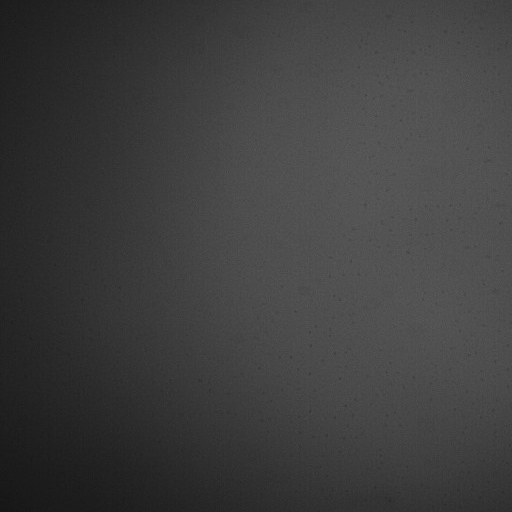}%
            \hfill
            \includegraphics[width=0.33\textwidth]{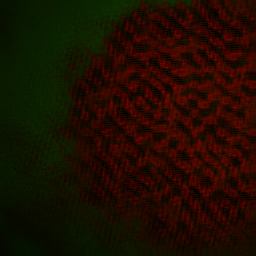}%
            \hfill
            \includegraphics[width=0.33\textwidth]{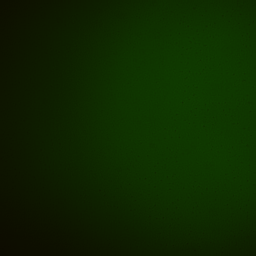}%
        \end{minipage}
        \caption{Failure example when two datasets have content differences. Left: single-channel patch of a blurry bright artifact with no biological content. Middle: Hallucinated translation when model is trained on a two-channel dataset that does not contain any similar artifacts. Right: Translation when patches with similar artifacts are added to the two-channel training dataset}
        \label{fig:failure}
\end{figure}

Our system relies on the assumption that the source and target datasets share a common content space. In cases where this assumption does not hold,  translation failures are likely.  In the example shown in \ref{fig:failure}, a single-channel multiplexed dataset contains large artifact regions (intense, cloud-like structures with no biological content) that are absent from the two-channel dataset. Without exposure to these patterns, the model produces hallucinated structures. In contrast, training on a dataset that includes similar artifacts enables the model to separate stains accurately. To address such failures, we propose identifying structures present in one dataset but missing from the other, and augmenting the underrepresented dataset through targeted sampling or synthetic data generation.



\end{document}